\documentclass[journal,11pt,twocolumn]{IEEEtran}

\usepackage{bm}
\usepackage{amsmath, amssymb,amsfonts,amsthm}
\usepackage{chngpage,float}
\usepackage{graphicx,epsfig}
\usepackage{multirow,booktabs}
\usepackage{stfloats}
\usepackage{graphicx}
\usepackage{subfigure}
\usepackage{multirow}
\usepackage{color}
\usepackage{epstopdf}
\usepackage{bbm}
\usepackage{booktabs}
\usepackage{algorithm,algorithmic,ragged2e,enumitem,url}
\usepackage{stfloats,cite}

\usepackage[colorinlistoftodos,textwidth=\marginparwidth]{todonotes}

\DeclareMathOperator*{\argmax}{argmax}

\newcommand{\bG}{\mathbf{G}}

\newcommand{\bd}{\bm{d}}

\newcommand{\bs}{\bm{s}}

\newcommand{\bfh}{\mathbf{h}}

\newcommand{\bfw}{\mathbf{w}}

\newcommand{\xv}{\mathbf{x}}
\newcommand{\yv}{\mathbf{y}}

\newtheorem{remark}{Remark}

\begin{document}

\title{Machine Learning in the Air}

\author{Deniz G\"{u}nd\"{u}z, Paul de Kerret, Nicholas D. Sidiropoulos, David Gesbert, \\ Chandra Murthy, Mihaela van der Schaar

\thanks{Deniz G\"{u}nd\"{u}z is with the Information Processing and Communications Laboratory, Department of Electrical and Electronics Engineering, Imperial College London, London, UK. Paul de Kerret and David Gesbert are with the Communication Systems Department, EURECOM, Sophia Antipolis, France. Nicholas D. Sidiropoulos is with the Department of Electrical and Computer Engineering, University of Virginia
Charlottesville, VA. C. Murthy is with the Department of Electrical and Computer Engineering at the Indian Institute of Science, Bangalore, India. Mihaela van der Schaar is with the University of California at Los Angeles (UCLA).}

\thanks{D. G\"{u}nd\"{u}z received support from European Research Council (ERC) under the European Union’s Horizon 2020 research and innovation program Starting Grant BEACON (grant agreement no. 677854). P. de Kerret and D. Gesbert are supported by the ERC under the European Union’s Horizon 2020 research and innovation program (agreement no. 670896). N.D. Sidiropoulos was partially supported by NSF CIF-1525194, ECCS-1608961, and ECCS-1807660. C. Murthy's work was supported in part by the Young Faculty Research Fellowship from the Ministry of Electronics and Information Technology, Government of India.}

}


\maketitle
\begin{abstract}
Thanks to the recent advances in processing speed and data acquisition and storage, machine learning (ML) is penetrating every facet of our lives, and transforming research in many areas in a fundamental manner. Wireless communications is another success story -- ubiquitous in our lives, from handheld devices to wearables, smart homes, and automobiles. 
While recent years have seen a flurry of research activity in exploiting ML tools for various wireless communication problems, the impact of these techniques in practical communication systems and standards is yet to be seen. In this paper, we review some of the major promises and challenges of ML in wireless communication systems, focusing mainly on the physical layer. We present some of the most striking recent accomplishments that ML techniques have achieved with respect to classical approaches, and point to promising research directions where ML is likely to make the biggest impact in the near future. We also highlight the complementary problem of designing physical layer techniques to enable distributed ML at the wireless network edge, which further emphasizes the need to understand and connect ML with fundamental concepts in wireless communications. 


\end{abstract}

\section{Introduction}\label{s:intro}

Recent advances in machine learning (ML) have caused a wave that has swept across all walks of science and engineering. The main premise of ML is to enable computers to learn and perform certain tasks (e.g., classification and prediction) without being explicitly programmed to do so. This is achieved by training algorithms on vast amounts of data available for the task to be accomplished. While the basic ideas and ambitions of ML go back to the 1950s, recent years have witnessed an unprecedented surge in interest in this area, fuelled by the availability of increasingly powerful computers, large and well-curated datasets, and developments in theoretical understanding of various learning algorithms.  

Arguably, the most impressive success stories of modern ML are due to the remarkable efficacy of deep learning, in the form of deep neural networks (DNNs), generative adversarial networks (GANs), and the resurgence of (deep) reinforcement learning ((D)RL) \cite{Goodfellow-et-al-2016, Sutton2018}. These tools have resulted in remarkable advances in audio and image recognition, natural language processing, recommender systems, and have beaten human grandmasters in chess and Go.  They have also led to many advances in applications from healthcare to autonomous driving, finance, marketing and robotics. The success of these approaches in many practical applications, and particularly the fact that they perform far better than disciplined approaches based on sound theory, has challenged the very foundation of our engineering education. Very few people believed that such a resurgence of `black box' methods would ever happen, much less that they would work so remarkably well in practice. Hence the latest wave in ML caught many communications engineers by surprise. But because we are engineers, we cannot look away from something that works. We have to understand it and use it to our advantage when possible. 
ML is certainly `in the air', with many special issues, workshops, special sessions and panels, exploring the potential and promises of ML for wireless systems. While the activity in this area is growing at an exponential rate, some seasoned researchers in the community are skeptical, and the impact of ML techniques in practical communication systems and standards is indeed yet to be seen.


Before we go into the challenges of applying ML in wireless systems, we would like to understand whether ML is really a novelty to communications researchers. Is it a completely new paradigm that can transform communications research and future communication systems, or is it yet another ``old wine in a new bottle'', presenting various old and known techniques with a new flavour? Indeed, the connections between ML and the theory of information transmission and storage are numerous and often striking. The fundamental problem of communication, as stated by Shannon \cite{Shannon}, ``reproducing at one point either exactly or approximately a message selected at another point,'' can in fact be recast as a classification problem. More specifically, symbol and sequence detection that constitute the core of any communication system are special cases of the general classification problem, which is at the heart of ML. Shannon's entropy, mutual information, and Kullback-Leibler divergence are widely used in ML as training objectives. Vector quantization (VQ) is key for source coding and rate-distortion, going back to Shannon \cite{Gersho:vecquant}. VQ is also known as k-means clustering -- a staple of unsupervised ML. Universal source coding inherently learns the distribution of the underlying information source in an online fashion \cite{LZ77, Willems:CWT}, and the most successful lossless compression algorithms are based on information theoretic principles, such as Lempel-Ziv and Burrows–Wheeler transforms, and have been successfully implemented for everyday use (gzip, pdf, GIF, etc.). Channel estimation is the task of learning a linear system in a supervised fashion when training/pilots are used. When (e.g., power amplifier) nonlinearities come into play, one has to learn a more general nonlinear system. Coding can be considered as controlled dimensionality expansion from the reduced-dimension {\em latent} information symbols to the channel input space, and decoding reduces things back to the original low-dimensional information space. 

Despite all the fascinating connections, there remain some key differences between generic ML and conventional wireless communications. Perhaps the most crucial differences are that i) in communications we have a fairly good grasp of what to expect by way of channel and system models, which obey physical laws; and ii) we have \emph{complete} control of what, when, and how to transmit. In principle this makes communications an overall better playing field for model-based solutions than generic ML applications.

The most striking aspect of the recent success of ML is its `data-driven' nature -- we have access to a lot of data nowadays, and hence, we can rely on data to draw conclusions and design systems like never before. The data-driven approach of ML is significantly different from the model-based approaches that have long dominated communication system design. Communication and networking engineers for many years have developed models with ever-increasing complexity and accuracy for the underlying physical communication channels, antenna patterns, data traffic, user mobility, interference, and many other aspects of communication systems. They have then designed highly complex modulation/ demodulation techniques, error correction codes, and networking protocols based on these models, which can be implemented efficiently (even on low-complexity and energy-limited mobile devices), and can enable reliable communications at fairly high data rates.  The model-based approach has been tremendously successful for communication system design, taking us from the first to the fifth generation (5G) of wireless networks, successfully keeping up with the rapidly growing demand for higher quality and lower latency content delivery. However, as we move towards implementing 5G networks and adopting a more flexible network architecture (network function virtualization, software defined networking, etc.), it is likely that there  will  be  many scenarios in which the  modeling assumptions used in traditional designs become questionable. For example, with network slicing and  multiple service classes, the interference can become highly non-stationary and non-Gaussian. Also, low-latency communications that should be supported with 5G may not allow accurate channel estimation, and short blocklength codes cannot benefit from the ergodicity of some of the randomness in the channel. Similarly, low latency requirements make the highly structured and modular layered network architecture highly suboptimal, requiring more integrated cross-layer designs, which increases the complexity of optimizing and operating these networks. These challenges point to the need for less structured solutions that are robust to model mismatches. Can the data-driven approach of ML be useful for designing such wireless communication systems and protocols? In this paper we will try to answer this question, focusing mainly on some exemplary applications of ML tools for lower layer design. We note here that the goal of this paper is not to provide a survey of recent results in this very active research area, but to highlight some of the striking recent results that promise significant gains compared to conventional physical layer design techniques, and provide a general discussion on why these techniques are promising, and the potential roadblocks for their implementation in real systems and adoption in standards. We refer the readers to excellent survey and overview papers on various aspects of ML in wireless communications to gather a more complete picture of its recent applications in different settings \cite{Hamed:survey, NN_Tutorial_CHen, ML_for_VN, Jagannath:Survey:IoT, Zhou_CR_ML, Simeone_TCCN_2018}. Next, we will go over some of the major challenges of applying ML in the lower layers of the protocol stack.



\subsection{Challenges of Applying ML Tools in Wireless Communications}


A major criticism for the data-driven approach to communication system design is the `black-box' nature of some of the ML algorithms, e.g., DNNs, and the lack of guarantees for performance; whereas communication engineers are accustomed to providing performance guarantees on error probability, interference level, channel outage, latency, etc. In many cases, such as emergency communication networks or for critical infrastructures, reliability and latency requirements can be extremely stringent. However, such provable guarantees hinge on the assumed channel, traffic, and device models, and their validity is as good as the accuracy of these models. Channel modeling, for example, no matter how ingenious, is always approximate, and the true channel dynamically evolves and is subject to all sorts of nonlinear / phase transition effects, from amplifier nonlinearities to loss of synchronization, which bring us closer to the realm of more general ML. On the contrary, the data-driven approach does not need powerful models, and instead can learn the optimal system architecture simply from available data. One particularly striking example is the use of the autoencoder as a general nonlinear detection mechanism -- without having to physically model, estimate, and explicitly implement an equalizer or an error control mechanism. The advantage of such an approach is that it can ``invert'' even unknown nonlinear channels directly, based only on training data and nothing else. Therefore, it is not clear which would provide a more reliable communication system: the one optimally designed based on complex yet approximate models, or another designed by black-box ML algorithms based on training data. While the former is limited by the accuracy of the model in representing the reality, the latter uses real data in the learning process, but the data is always limited in size and generalizability. We expect that ML-based solutions will be effective particularly when an accurate model for the problem of interest is not available (i.e.,`model-deficit' as referred to in \cite{Simeone_TCCN_2018}), and a sufficiently large and representative training dataset is available.  

The `black box' aspect of DNNs also brings along the \textit{interpretability} problem. Understanding the reasons behind the success or failure of ML methods, particularly those based on DNNs, is an on-going research challenge \cite{interpret:arxiv:17}, which is yet to be addressed satisfactorily. From an engineering perspective, not knowing the reasons behind the decisions taken by an algorithm makes it very difficult to tackle failures, or to predict the impact of changes in the environment on the performance. Also relevant for communication networks is to guarantee some sense of fairness across users, such that they are not penalized unintentionally by a ML algorithm due to the type of their device, their location, protocol being used, etc.

Another challenge of applying data-driven ML tools to wireless systems is the limited availability of training data. Unlike in computer vision, speech processing, or health-care applications, in most wireless applications standardized datasets for testing and comparison of proposed ML techniques are not available. However, we expect that, with the increasing adoption of ML techniques, more public datasets will become available to the community. There are already a few datasets that can be used to perform and compare some basic ML tasks on wireless signals \cite{OShea:GNU:16, Alkhateeb2019, Brazil:dataset, Arnold:dataset}. 

Even if such datasets become available, it is questionable whether success on such datasets can promise success in other channel and network conditions. Wireless channels are often highly non-stationary, and offline training on a generic dataset may not lead to satisfactory results when tested on a very different wireless environment. This may require online training of the existing models to adapt them to the current scenario; however, training time of an ML model for reasonable performance is often beyond the operation timescales of communication systems. 

In order to deal with limited training data sizes, and under the condition that there exist models (e.g., for the channels) that are reasonably accurate, one can generate synthetic data from the model, which can be used for offline training. In this case, by a judicious choice of the architecture, one can arrive at an ML algorithm that (at least empirically) outperforms its conventional counterparts, when algorithms of similar computational complexity are compared. This points to another setting in which ML-based techniques have proven useful: even when the system model is accurately or perfectly known, the optimal solution may be too complex, or even intractable \cite{LearnWMMSE} (referred to as `algorithm-deficit' in \cite{Simeone_TCCN_2018}). In such a case, the model can be used to generate data, which can then be used to train a limited-complexity ML-model, which can either try to imitate an available model-based approximate or optimal solution, or directly achieve the optimal performance. Such an approach has been shown to provide  approximate solutions to even NP-hard problems using moderate computational resources \cite{Vinyals:NIPS:15, NPhard:app}, or to outperform human experts in fully known yet highly complex models, such as chess, Go, or Atari games \cite{Atari:DL, Silver:AGoZ}. 


Another issue raised  when adopting ML-based techniques in wireless communication networks is the limited computational and memory resources available to most wireless devices, especially low-complexity terminals at the network's edge. Many of the impressive results with data-driven ML techniques are obtained using very powerful computing machinery and massive datasets, which may not be possible to reach by mobile devices with limited computation, memory, and energy resources. The computing power of even the most recent mobile devices is orders of magnitude less than high performance computers used to train complex ML models. Also, each wireless end-user device typically has only a limited amount of data, further limiting the training capabilities. The current approach to overcome the limitations of wireless devices is cloud or edge processing, in which all the data available at wireless devices are transferred to an edge or cloud unit, where a powerful ML algorithm can be centrally trained using all the data. However, such a solution comes at the cost of transferring the data from energy and bandwidth limited wireless edge devices to a central edge or cloud processor, and the latency this would incur -- not to mention privacy concerns, which are increasingly becoming a serious challenge to centralized data processing. This necessitates new ways of achieving decentralized learning in the wireless setting. Decentralized learning and decision making is ultimately limited by how much information is let to flow between the learning devices and how much noise corrupts the local device's information. Clearly, algorithms which can adapt to arbitrarily distributed information settings would be highly desirable.


In the rest of this paper, we will highlight some specific problems in wireless communication networks, and in particular at the physical layer, where we believe ML techniques can make a significant impact. In some of the settings, data-driven solutions are posed to solve hard wireless networking problems, trained on data generated from existing models. These emphasize the use of ML as an optimization technique to obtain solutions that can surpass the state-of-the-art. We will also highlight some applications in which data-driven ML techniques are suitable due to the lack of accurate models. Although we provide pointers to a large number of key references, the presented examples are naturally influenced by our personal research experiences and interests. Nonetheless, we believe that the highlighted observations are likely to  `generalize' to other relevant problems and scenarios.


\section{Deep Learning Based Detection and Decoding}

Data detection over a noisy channel, which is an essential component of any communication system, is inherently a classification problem. Current detection systems rely on model-based solutions, employing a mathematical model describing the underlying communication channel. Moreover, we typically use a detector derived assuming perfect channel state information (CSI) at the receiver, with the channel state replaced by its estimate computed from training symbols. This renders the detector sub-optimal in the presence of CSI estimation errors, and a well-trained ML algorithm can outperform classical approaches. In the context of data detection under a Poisson channel model (which arises in molecular communication),
in \cite{Farsad_TSP_2018}, 
a recurrent NN (RNN) is used in the presence of intersymbol interference (ISI). While the proposed RNN structure can be trained to learn to disentangle the impact of ISI without any additional information, the performance of the classical Viterbi decoder (VD) depends heavily on the accuracy of CSI as well as the memory length of ISI in the channel. In \cite{Farsad_TSP_2018}, authors also train a detector based purely on data collected from a molecular communication channel. Since accurate models for this system are lacking, they show that, under CSI estimation errors, the NN-based detector performs significantly better than state-of-the-art detectors. This result corresponds to a fully data-driven approach for a complex system with hard-to-model imperfections and nonlinearities, where ML provides an attractive alternative. 


The detection problem is also studied in \cite{Samuel2017} considering a MIMO channel:
\begin{align}
    \mathbf{y} = \mathbf{H}\mathbf{x} + \mathbf{w},
\end{align}
where $\mathbf{y} \in \mathbb{R}^N$ is the received vector, $\mathbf{H}\in \mathbb{R}^{N\times K}$ is the channel matrix, $\mathbf{x} \in \{-1, +1\}^K$ is the unknown channel input vector consisting of independent and equally likely binary symbols, and $\mathbf{w}\in \mathbb{R}^{N}$ is the noise vector consisting of independent zero-mean Gaussian variables of unknown variance. Even under the perfect knowledge of the channel model and the channel matrix $\mathbf{H}$, as the dimensionality of the problem, $N \times K$, increases, the optimal maximum likelihood (ML) detector becomes impractical due to formidable computational complexity. The authors propose using a DNN-based detector. The challenge here is to find the best way to feed the CSI to the DNN to allow the network to learn to exploit this additional information. The authors exploit the structure of a projected gradient solution, and design the network architecture accordingly. In particular, they feed into each layer of the network $\mathbf{H}^T \mathbf{y}$, $\mathbf{v}_k$, $\mathbf{x}_k$ and $\mathbf{H}^T\mathbf{H} \mathbf{x}_k$, where $\mathbf{v}_k$ and $\mathbf{x}_k$ are obtained iteratively as outputs of each layer of the DNN. The results in \cite{Samuel2017} show that the DNN-based decoder achieves comparable performance with a high-complexity decoder based on semi-definite relaxation, while running 30 times faster. This is a good example of a solution that combines DNN-based ``black-box'' solution with domain knowledge, which is used to ``steer'' the DNN to exploit the available information in an efficient manner. Note that, the decoder is not provided any additional information, and in theory should be able to learn to mimic this structure; however, providing the same information in the most convenient form can speed up the learning process, and avoid suboptimal local optima. 


In \cite{Ye_WComLet_2018}, the authors study the data detection problem over a known channel model, without CSI at the receiver. The DNN-based decoder in this case is trained to output the estimated data symbols based purely on the received signal, without explicitly estimating the wireless channel. The DNN is trained using synthetically generated input-output data. Specifically, the channel is drawn from the wireless world initiative for new radio (WINNER II) model, which models a typical urban channel with 24 sample-spaced paths. The DNN consists of an input layer, three hidden layers and an output layer, with a heuristically selected number of neurons in each layer. The Relu function is used as the activation function in all but the output layer, where the sigmoid function is used to map the output to the interval $[0,1]$. The mean squared error between the transmitted and predicted symbols is used as the loss function for training the DNN. Numerical results illustrate several interesting points. First, when sufficiently many pilots are present, the bit error rate (BER) performance of the DNN-based decoder matches that of the MMSE receiver. However, under non-ideal conditions, such as fewer pilots, absence of the cyclic prefix, or nonlinear clipping noise, the DNN-based decoder can significantly outperform the MMSE receiver.  Of course, the MMSE receiver is no longer optimal under these non-idealities, and another hand-crafted solution that address them could perform as well or better than the DNN-based decoder. Nonetheless, the deep learning approach offers a relatively straightforward and promising solution, that can potentially deal with a variety of non-idealities in a robust manner.  In \cite{Jiang_ArXiv_2018}, the authors carry this idea further and illustrate over-the-air results of an online trainable OFDM receiver.

\subsection{Channel Estimation}

Recall that MMSE channel estimation entails knowledge of the channel statistics and a potentially computationally expensive conditional mean computation. In \cite{Neumann2018LearningTM}, the authors model the channel as conditionally Gaussian distributed given a set of (hyper)parameters. These hyperparameters are also random, whose distribution is eventually learned from training data. Now, it turns out that the MMSE estimator under this model can be written as a linear estimator, with the weights depending on the statistics of the hyperparameters. By vectorizing the MMSE estimate, the authors write the estimator in a form that is amenable to implementation as a feed-forward neural network with two linear layers connected by a nonlinear activation function. These layers are made learnable, and are trained via stochastic gradient descent with the mean squared channel estimation error as the loss function. It is shown that, under certain assumptions, this can lead to a computationally inexpensive, near-optimal MMSE estimator when the channel covariance matrix are Toeplitz and have a shift-invariance structure. Simulation results suggest that the NN based channel estimator outperforms state-of-the-art estimators and has low complexity. 

In the context of wideband channels, \cite{Soltani_ArXiv_2019} models the channel time-frequency response as an image, and the pilot based samples as a low-resolution sampled version of the image. The authors use convolutional neural networks (CNN) based image super-resolution and image restoration techniques, implemented using a three-layer and a $20$ layer CNN, to estimate the channel, with the mean squared error as the loss function. Empirically, the performance is demonstrated to be similar to that of an ideal MMSE estimator that has perfect knowledge of the channel statistics. 

As a final note on DNN-based channel estimation, we note that the above ideas have been extended to the case where the receiver has fewer RF chains than antenna elements, for example, in mmWave systems. In this case, the key challenge is that the receiver has to estimate the channel from compressed measurements. In \cite{Koller2018MachineLF, He_ArXiv_2018}, it is shown that these estimators can even outperform estimators based on sparse signal recovery, when trained with sufficient amount of data. 

\subsection{Learning to Decode}

While the above works mainly focus on detecting the channel input symbols, DNNs can also be used to recover coded symbols. Decoding of codewords from a certain channel code is another classification problem. However, the number of classes to classify the received signal into grows exponentially with the blocklength, leading to exponentially growing training complexity. Therefore, most of the current approaches to DNN-based channel decoding incorporate DNNs into the existing decoder structures. For example, \cite{Nachmani2016} uses a NN to learn the weights that should be assigned to the Tanner graph of the belief propagation (BP) algorithm. In \cite{Cammerer2017}, authors propose improving the performance of conventional iterative decoding for polar codes by complementing it with NN-based components. In particular, they divide the decoder into subblocks, each of which is replaced by a NN-based decoder, and the results of these decoders are fed into a belief propagation (BP) decoder. This allows controlling the training complexity by adjusting the number of subblocks. 
A fully DNN-based channel decoder is considered in \cite{Gruber2017}. To keep the complexity reasonable, codelength is limited to $16$ while the code rate is fixed to $1/2$. The authors trained the decoder NN both for a polar code and a random code. While a performance close to a maximum aposteriori (MAP) decoder can be achieved for the polar code, the gap to the MAP decoder performance is much larger for the random code. Although this gap can be closed with increasing the number of training epochs, the result highlights the point that NNs are most effective when the data has an underlying structure that can be learned. The authors also considered limiting the set of codewords observed during training. This is to test whether the NN-based decoder can generalize to unseen codewords. They observed that this was indeed the case for the polar code; the decoder was able to learn to decode codewords it has never seen before, which can only be explained by the fact that the NN-based decoder has learned the structure of the decoding algorithm. This is not the case for the random code, which did not have any particular structure that could be exploited by the NN-based decoder.


\subsection{Observations}

Certain features that are common to the aforementioned works are worth mentioning. First, most of them use the so-called one-hot representation of the transmitted signal \cite{Felix_ArXiv_2018, Raj_ComLet_2018} . In the one-hot representation, the signal is represented as a binary vector of length equal to the number of possible signals. The binary vector contains a single $1$ at the location corresponding to the transmitted signal, and zeros everywhere. While one-hot encoding typically provides better results as it prevents any numerical ordering between the inputs, it also leads to an exponentially growing input size for channel decoding.  

The output layer of the DNN that attempts to reconstruct the input signal is typically chosen as the sigmoid function. In this case, the DNN attempts to output the likelihoods of possible signals, which is useful, for example, in detection of coded symbols, where the bit log likelihood ratios need to be fed into a channel decoder. 

Another interesting recent development is the use of recurrent neural networks with long-short term memory (LSTM), which allows for smaller generalization error \cite{Farsad_TSP_2018}. This allows for unseen channel instantiations to be handled effectively. 

\section{Autoencoders for End-to-end Communication System Design}

An \textit{autoencoder} is a pair of NNs, called the \textit{encoder} and the \textit{decoder} networks, that try to learn the identity function; that is, they try to recover the input at the output. It is an unsupervised learning technique as it does not require any labels. The output of the encoder network, called the \textit{bottleneck layer}, is the input to the decoder network. Typically the bottleneck layer has lower dimension than the input data, and if the autoencoder can learn to recover the input reliably, this means that the bottleneck layer carries all the information necessary to reconstruct the input; and hence, is a compressed version of the input signal. Autoencoders are typically used in machine learning for feature extraction or as generative models for data \cite{Goodfellow-et-al-2016}.

The correspondence of the autoencoder structure to a communication system with an encoder and a decoder is quite obvious. Indeed, autoencoders have been successfully applied to image \cite{Setiono:autoencoder, Balle:ICLR:17} and video \cite{Han:deepVideo} compression, which can be considered as communication over a finite-rate error-free channel. End-to-end learning of encoder and decoder functions for communications over a physical layer channel is first proposed in \cite{Oshea:16:arXiv}, and later expanded in \cite{Oshea2017}. The noisy communication channel that connects the output of the encoder NN to the input of the decoder NN is treated as an untrainable layer with a fixed transformation. This end-to-end training of the physical layer bypasses the modular structure of conventional communication systems that consists of separate blocks for data compression, channel coding, modulation, channel estimation and equalization, each of which can be individually optimized. While this modular structure has advantages in terms of complexity and ease of practical implementation, it is known to be suboptimal. An autoencoder is trained for coding and modulation over an additive white Gaussian noise channel in \cite{Oshea2017}, and it is shown to have a performance very close to conventional coding and modulation scheme in short blocklengths.

The aforementioned works on autoencoder-based end-to-end physical layer design assume a known channel model, and the encoder and decoder networks are trained jointly by simulating many realizations of this channel model. While channel models for wireless channels are considered to be accurate in general, they may still have mismatch with the real channel experienced by the transceivers, limiting the overall performance of the system. An alternative would be to use a GAN architecture to learn a channel model based on real data collected from the channel. This can provide a more accurate model of the channel, particularly if sufficient data can be collected from the channel. In \cite{Ye_ArXiv_2018}, the authors propose to use the learned GAN as the channel layer between the encoder and decoder NNs of an end-to-end communication system. 

A fundamental challenge in training autoencoders directly on a real channel is the significant delay this may cause. Since the encoder and decoder must be trained jointly, the backpropagation has to propagate the gradient from the receiver to the transmitter, requiring a feedback link during training, which would significantly slow down training. To circumvent this limitation, \cite{Dorner2017} proposes a two-phase training approach: the first phase uses a channel model as before and the encoder and decoder are trained based on this model as before. Once these networks are deployed at the transmitter and the receiver, the receiver network is trained further based on the transmission of known signals from the transmitter. This is similar to pilot transmission in channel estimation, and does not require feedback to the transmitter. 


\begin{figure}[t]
	\begin{center}
		\includegraphics[width = 0.5\textwidth]{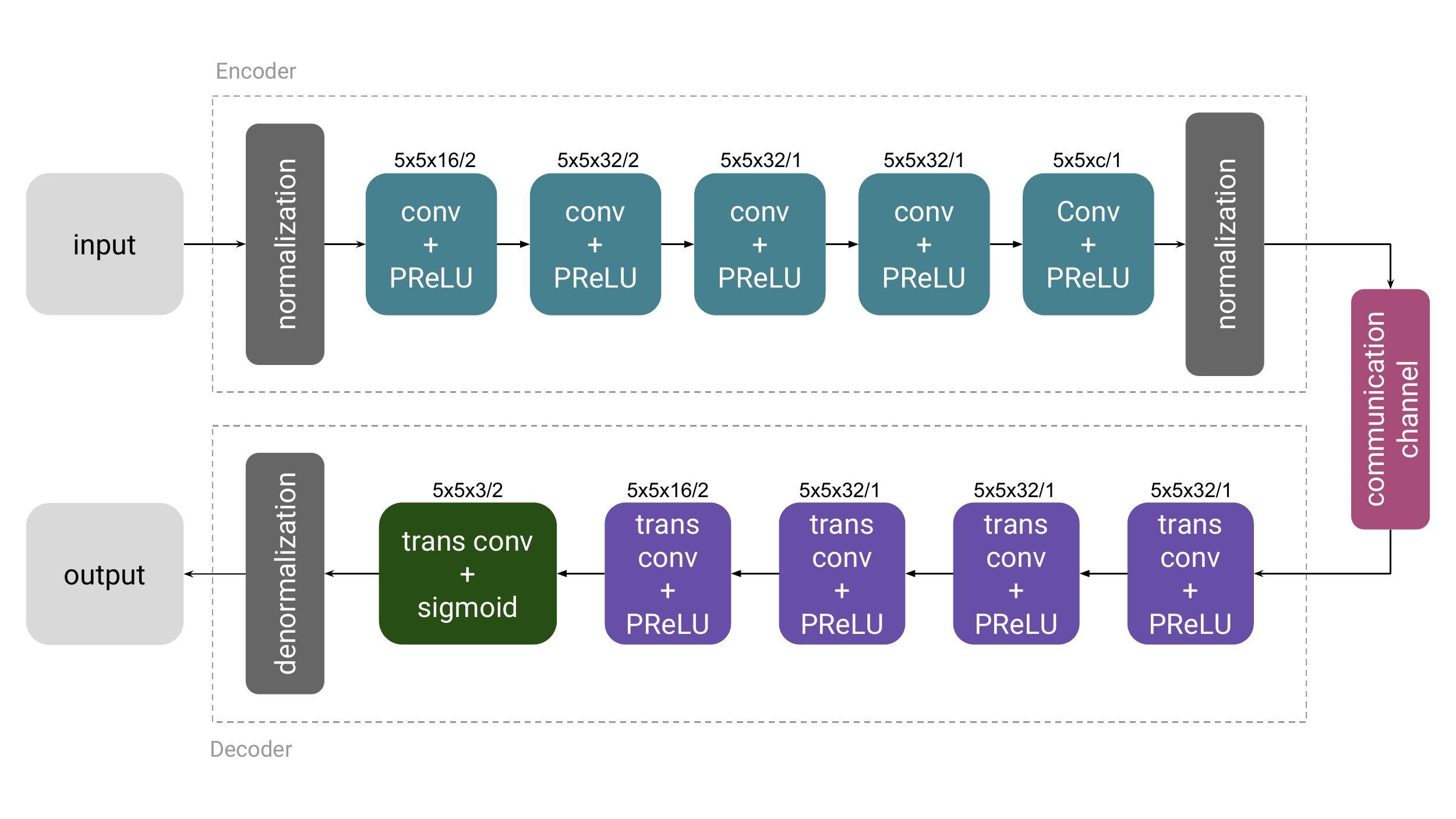}
	\end{center}
		\caption{Encoder and decoder NN architectures used in the implementation of the deep JSCC scheme in \cite{Eirina:deepJSCC}.} \label{fig:architecture}
\end{figure}

\subsection{Joint source-channel coding (JSCC)}

All the above works have exclusively focused on transmitting bits over the noisy channel, that is, the goal is to design error correction codes jointly with modulation, channel estimation, etc. Note that when the input to the encoder is a bit sequence, there is no structure in the data, and the goal is to learn the best mapping of the message bits into the channel input space, and jointly the best inverse mapping. This is achieved mainly by distributing the input signals as much as possible in the channel input space within the constraints of the transmitter, and taking into account the random channel transformation. However, in many real applications, the goal is to transmit some information signal, e.g., a picture, video, or an audio signal, which is not in the form of a sequence of equally likely bits, and typically has significant redundancy. 

\begin{figure}[t]
	\begin{center}
 		{\includegraphics[width=0.5 \textwidth]{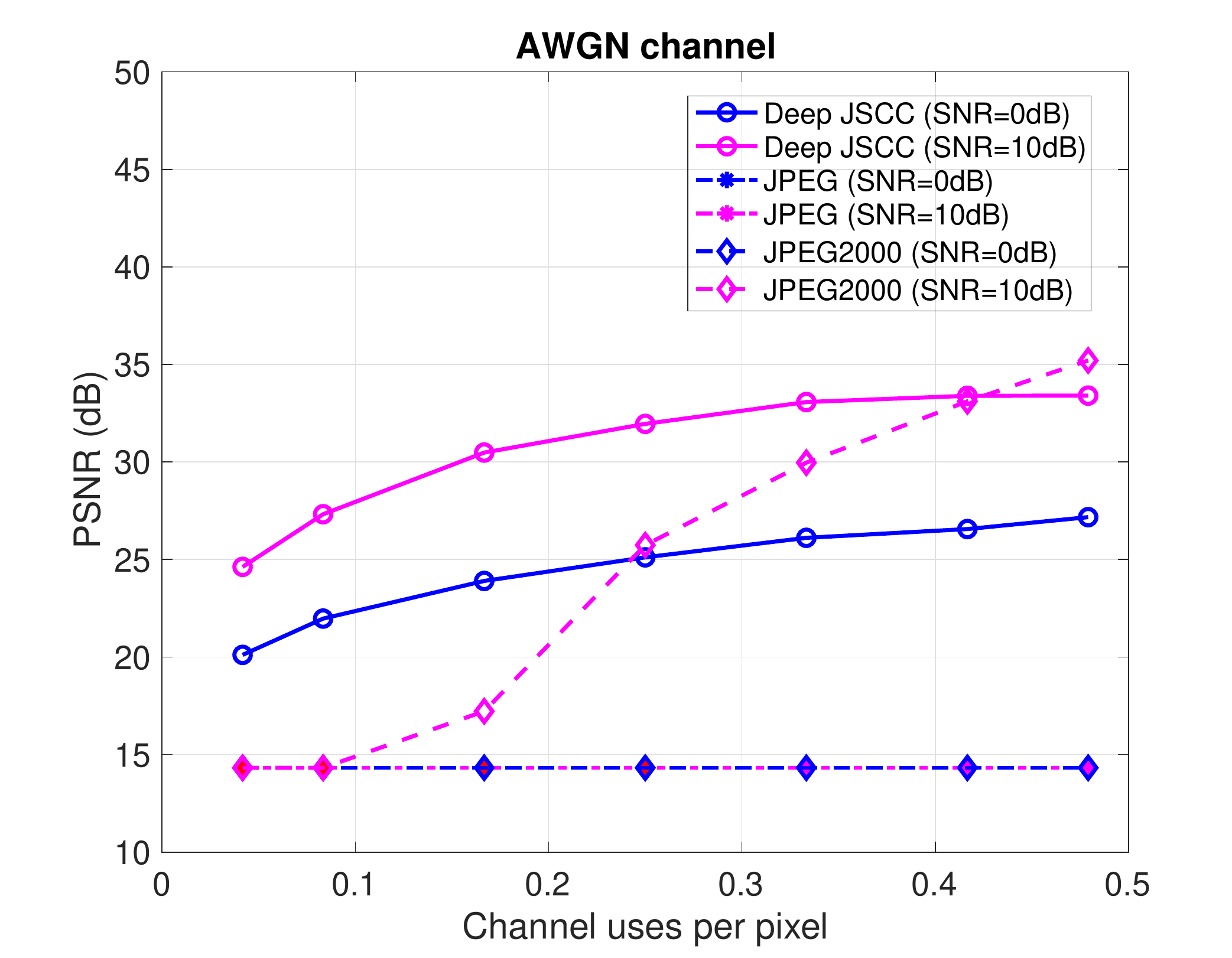}} 
	\end{center}
		\caption{Performance of the deep JSCC algorithm in \cite{Eirina:deepJSCC} on CIFAR-10 test images transmitted over an AWGN channel with respect to the available channel bandwidth per image pixel for different $\mathrm{SNR}$ values. For each case, the same $\mathrm{SNR}$ value is used in training and evaluation, and a different network is used to obtain each point in the curve.} \label{fig:cifar10_awgn_PSNR_vs_CR}
\end{figure}

The current standard approach to transmission of such signals is to first compress them with a source coding algorithm in order to get rid of the inherent redundancy, and to reduce the amount of transferred information; then the compressed bitstream is encoded and modulated over the channel. Shannon's \textit{separation theorem} proves that this two-step separate source and channel coding approach is optimal theoretically in the asymptotic limit of infinitely long source and channel blocks \cite{Cover2006}. However, in practical applications, JSCC is known to outperform the separate approach, particularly in short blocklength and low SNR regimes. Many emerging applications from the Internet-of-things (IoT) to autonomous driving and to tactile Internet require transmission of high data rate information (image/video, various sensor measurements) under extreme latency, bandwidth and/or energy constraints, which preclude computationally demanding long-blocklength source and channel coding techniques. However, characterizing the optimal JSCC in non-asymptotic regimes has remained an open problem, even for fully known source and channel distributions, and it is significantly more challenging for the transmission of complicated sources, such as images or videos, for which we do not have good statistical models.

\begin{figure}[t]
	\begin{center}
 	{\includegraphics[width=0.53\textwidth]{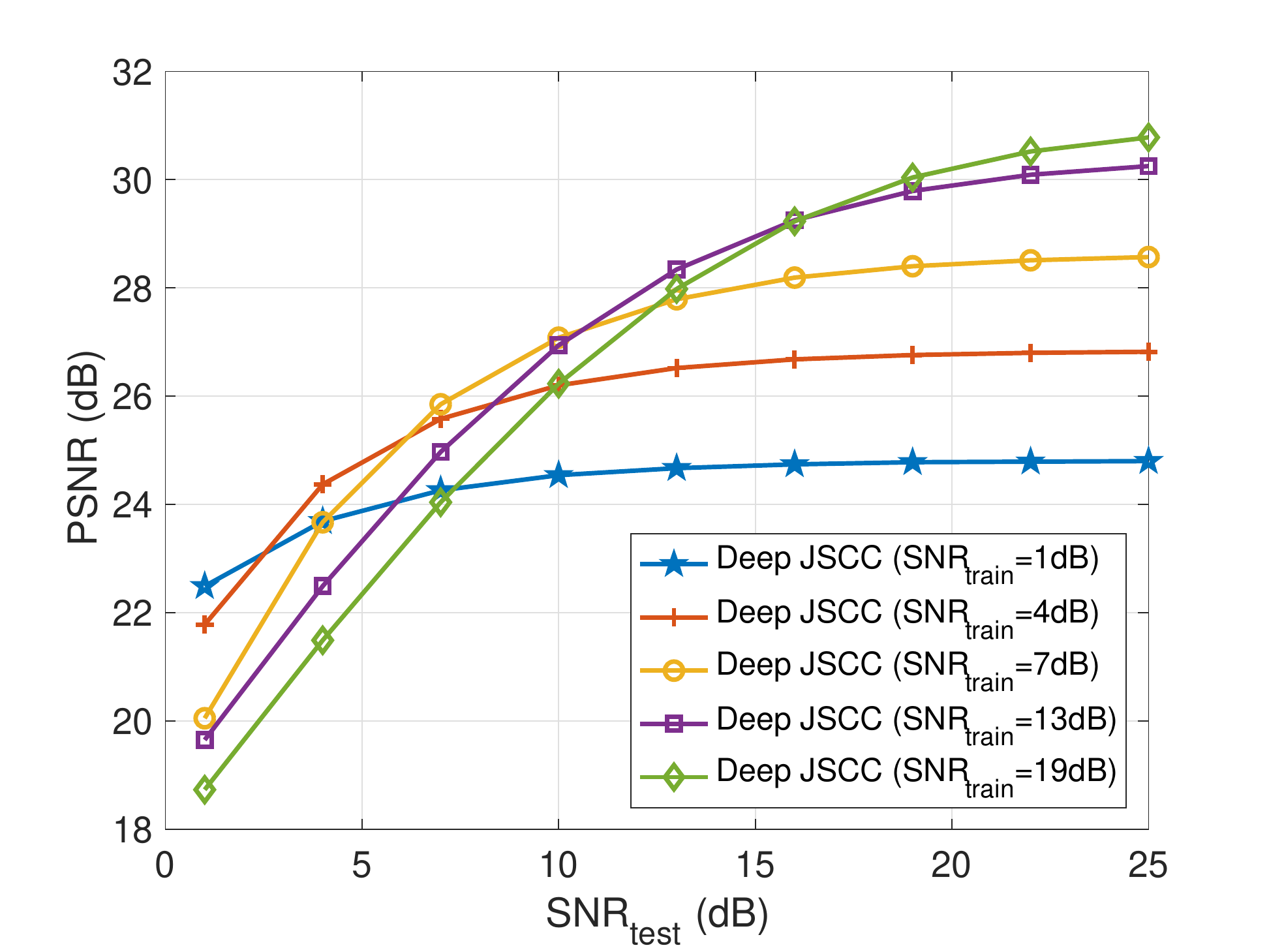}}
	\end{center}
		\caption{Performance of the deep JSCC algorithm  in \cite{Eirina:deepJSCC} on CIFAR-10 test images with respect to the channel SNR over an AWGN channel for a bandwidth compression ratio of $1/12$, i.e., 1 channel use per 12 pixels. Each curve is obtained by training the encoder/decoder network for the indicated $\mathrm{SNR_{train}}$ value.}
	\label{fig:cifar10_awgn_1over12}
\end{figure}

Alternatively, a deep JSCC architecture can be trained to map the underlying signal samples directly to channel inputs. Such an architecture is studied for transmission of images over wireless channels in \cite{Eirina:deepJSCC}. This can be considered as an ``analog'' JSCC scheme since, unlike digital systems built upon the separation approach, the input signal is never converted into bits, and the channel input signal is not limited to a finite number of constellation points. The deep JSCC architecture proposed in \cite{Eirina:deepJSCC} is illustrated in Fig. \ref{fig:architecture}. This fully convolutional architecture allows compression of images of any size. The results illustrated in Fig. \ref{fig:cifar10_awgn_PSNR_vs_CR} show that deep JSCC outperforms state of the art digital image transmission schemes, e.g., JPEG or JPEG2000 image compression followed by capacity-achieving channel codes, particularly in the low SNR and short channel bandwidth regimes. Note that, both JPEG and JPEG2000 fail completely at $\mathrm{SNR}=0$~dB. For $\mathrm{SNR}=10$~dB, JPEG2000 can provide reasonable quality if the channel bandwidth is sufficiently large. A few aspects of deep JSCC are particularly worth mentioning: First of all, it provides non-trivial image reconstruction even at very low SNR values and limited channel bandwidths, i.e., in the case of short blocklengths. Moreover, thanks to the analog nature of the encoder, the performance also behaves like analog modulation schemes, and exhibits graceful degradation with channel SNR. This can be observed in the performance curves in Fig. \ref{fig:cifar10_awgn_1over12}. A deep JSCC architecture trained for a particular target channel SNR value gracefully degrades if the channel SNR falls below this value, and its performance improves if the channel SNR goes above the target value. 

This analog behaviour is particularly attractive for broadcasting to multiple receivers, or when transmitting over a time-varying channel. Indeed, it is shown in \cite{Eirina:deepJSCC} that the performance improvement of deep JSCC compared to conventional digital schemes is much higher over fading channels. Note also that, while learning channel codes is challenging even for very limited blocklengths, deep JSCC can achieve performance levels above or comparable with state-of-the-art digital techniques even over large blocklengths.

It is shown in \cite{Kurka:arxiv:19} that this deep JSCC architecture also allows bandwidth adaptation through successive refinement; that is, an image can be transmitted over $n$ layers, and a user receiving the first $k$ layers can recover the image with peak signal-to-noise ratio $\mathrm{PSNR}_k$, $k=1, \ldots, n$. While $\mathrm{PSNR}_1 < \cdots < \mathrm{PSNR}_n$ as expected, $\mathrm{PSNR}_k$ is very close to the performance one would obtain if the image was transmitted targeting the total bandwidth available for the first $k$ layers; that is, transmitting the image in layers comes at almost no additional cost, providing seamless bandwidth adaptivity.

\section{Machine Learning Based Resource Allocation}


An important class of problems where modern ML techniques can help is formed by (NP-) hard resource allocation and decision / scheduling problems -- which are very common in wireless communications and networking. Examples range from classical multi-user detection to sum-rate optimal power control, multi-user scheduling, and transmission control -- or ``smart'' data-driven TCP-IP. In the following paragraphs, we will review some illustrating examples.

The case of joint multicast beamforming and antenna selection is considered in \cite{LearnAntSel}, where it is shown how a DNN can be used to successfully solve the discrete optimization part of the problem. This is an example of a hybrid strategy, where a DNN is employed to solve part of the problem, synergistically with classical optimization. 

Another example of this synergy is recently demonstrated in the context of power system state estimation. This is another nonconvex quadratically constrained quadratic programming problem that is typically tackled using Gauss-Newton-type approaches, which are however sensitive to initialization. Gauss-Newton algorithm converges very fast when initialized in the neighborhood of the true state, but may fail to converge, or even diverge, if the initialization is poorly chosen. The idea of \cite{LearnDSSE} is to train a neural network to {\em learn to initialize}.  This entails a special design of the ($\epsilon$-insensitive) learning function to {\em lower the bar} in a way, and focus just on getting in the right ball park of the optimal solution. Lowering the bar in this way has important advantages in terms of the required neural network complexity, and accordingly the required sample complexity. 

Beamforming for minimum outage \cite{BMF4MinOut} has also been proven to be NP-hard even when the channel distribution is known {\em exactly}, and in fact no practically good approximation algorithm was known until very recently. Yet, relying on a sample average `counting' approximation of outage, simple smoothing, and stochastic gradient updates, a lightweight and very effective algorithm was recently designed \cite{Learn2BMF} that performs remarkably well, using only recent channel data and nothing else. The problem is formulated as follows: 
\begin{equation} \label{cost_Probability}
\min_{\bfw \in \mathcal{W}} \biggl\{ F(\bfw) := \textrm{Pr}\biggl(|\bfw^H\bfh|^2 < \gamma\biggr)  \biggr \},
\end{equation}
where $\gamma > 0$ denotes the outage threshold and $\mathcal{W} \subset \mathbb{C}^N$  is a simple (element-wise or sum) power constraint. 
We can equivalently express \eqref{cost_Probability} as
\begin{equation}\label{eq:SO}
\min_{\bfw \in \mathcal{W}} \textrm{Pr}\biggl(|\bfw^H\bfh|^2 < \gamma\biggr) \Leftrightarrow
\min_{\bfw \in \mathcal{W}}\mathbb{E}_{\bfh} [ \mathbbm{1}_{\{|\bfw^H \bfh|^2 < \gamma\}}].
\end{equation}
Define
\begin{equation}
f(\bfw; \bfh):= \mathbbm{1}_{\{|\bfw^H \bfh|^2 < \gamma\}} = \left\{
\begin{aligned}
1, &   ~\text{if}~ |\bfw^H \bfh|^2 < \gamma \\
0,  &  ~\text{otherwise}
\end{aligned}
\right.
\end{equation}
as the indicator function of the event $|\bfw^H \bfh|^2 < \gamma $. Consider a given set of `recent' channel realizations $\mathcal{H}_T := \{\bfh_t\}_{t=1}^T$. Utilizing $\mathcal{H}_T$, we may construct the following sample average estimate of $\mathbb{E}_{\bfh}[ f(\bfw; \bfh)]$
\begin{equation}
\hat{F}(\bfw;\mathcal{H}_T) := \frac{1}{T} \sum_{t=1}^{T} f(\bfw; \bfh_t).
\end{equation}
The interpretation is that we  minimize the total number of outages over (`recent') channel history - very reasonable, since under 
appropriate mixing conditions we have
\begin{equation}
  \lim_{T \rightarrow \infty} \hat{F}(\bfw;\mathcal{H}_T) = \mathbb{E}_{\bfh}[ f(\bfw; \bfh)] = F(\bfw), \forall \; \bfw \in \mathcal{W},
\end{equation}
almost surely. Replacing $F(\bfw)$ by $\hat{F}(\bfw;\mathcal{H}_T)$ in \eqref{cost_Probability}, we obtain 
\begin{equation} \label{Samples_approximation}
\min_{\bfw \in \mathcal{W}} \hat{F}(\bfw;\mathcal{H}_T).
\end{equation}
The final step is to construct a smooth approximation of $f(\bfw;\bfh)$, and optimize the resulting function using stochastic gradient descent. The details can be found in \cite{Learn2BMF}, which also shows that the approach works unexpectedly well, on a problem that has challenged many disciplined optimization experts for years. 

Finally, the sum-rate optimal power control problem is known to be NP-hard, but we have good, albeit computationally expensive, approximation schemes at our disposal. These include the iterative weighted minimum mean squared error (WMMSE) approach\cite{Christensen2008,Shi2011}, and successive convex approximation \cite{Kaleva2015}. These algorithms are too complex for practical implementation, but a key idea advocated in \cite{LearnWMMSE, LearnWMMSE-SPAWC} is that we can take this complexity offline by training a DNN to mimic the input-output behavior of the WMMSE algorithm. The way to do this is to use historical (measured) and/or simulated channel data, run the WMMSE algorithm offline to generate the associated power allocation values, and use these input-output pairs to train the DNN. At run time, we simply pass the input through the trained neural network, which is far cheaper than running WMMSE online. The remarkable thing is that this works well. 

The approach can be further refined by training the network to optimize sum rate directly as described in \cite{Lee2018}. In that case, the sum rate is directly differentiated with respect to the coefficients of the DNNs, which allows to further improve the performance. The existing state-of-the-art solutions and the approximation approach described above are still used to initialize the optimization, and hence avoid the inefficient local optima.

\subsection{Machine Learning for Decentralized Communication Design}
While the resolution of the resource allocation problems above exploit a central common intelligence, many networking problems require decentralized optimization. Such settings include, for instance, coordination and cooperation tasks among radio devices in the absence of a central controller. Such problems can be linked to so-called   \emph{team decision (TD)} and \emph{decentralized control} problems, which are notoriously difficult to tackle. In TD problems, multiple-agents aim at cooperating to achieve a common goal on the basis of different information. The multi-agent setting is rendered particularly challenging due to the fact that local input data at each agent is typically noisy (with the noise level possibly being different from agent to agent) and the communication capability for information exchange between the agents is often limited in terms of rate and/or latency.

The derivation of robust multi-device decision-making algorithms with arbitrary input uncertainties across agents is well known to be a very challenging task which does not find a solution via conventional optimization methods. Team decision problems were first formulated by Radner in \cite{Radner1962} and then studied by Marschak in \cite{Marschak1972}. Although some particular simple cases could be solved (e.g., a linear objective), the general problem remains open, with no good approximate solution. This makes this class of communication design problems an interesting playing field for ML. 


\subsubsection{Problem Formulation}

Communication design approaches involving multiple radio devices optimizing some communication, physical layer or resource allocation parameters jointly for the sake of maximizing the network performance can be recast as multi-agent coordination problems. Typically, the agents (aka radio devices) need to optimize their transmission decision on the basis of some noisy channel state information which represents the input data at the agent. 

Consider a multiple-agent setting with $n$~agents where agent~$j$ takes a decision $\bd_j$ using the information locally available, denoted by $\yv_j$:
\begin{equation}
\bd_j=\bs_j(\yv_j),
\end{equation}
where $\bs_j$ can be any arbitrary function from the information space to the decision space. Information~$\yv_j$ available at agent~$j$ is the result of sensing, estimation, feedback, and potentially any information sharing through the backhaul network occurring before the actual transmission. Due to the limited resources available, and the inevitable delay introduced by this process, the resulting estimate obtained is expected to be a potentially imperfect and/or incomplete estimate of the true representation of the world~$\xv$. 

We focus here on the fully cooperative scenario, in which all the agents aim at jointly maximizing a common utility function~$u$ in an expected sense. The optimization problem can then be written as
\begin{equation}
\left ( \bs_1^{\star},\ldots, \bs_n^{\star}\right)=\argmax_{\bs_1,\ldots,\bs_n}  \;\;\mathrm{E}\left[ u(\xv,\bs_1(\yv_1),\ldots,\bs_n(\yv_n)) \right] 
\label{eq:TD_1}
\end{equation}   
with $\xv,\yv_1,\ldots,\yv_K$ jointly distributed according to~$p_{\xv,\yv_1,\ldots,\yv_K}$. Importantly, we assume that all the agents have access to the knowledge of this distribution, or equivalently, as it will become clear later on, that the training database is \emph{commonly} known at all agents. 

The optimization problem \eqref{eq:TD_1} is a functional optimization problem, with the decentralized aspect being taken into account by the functional dependencies of the decision functions and the joint distribution between the estimates~$p_{\xv,\yv_1,\ldots,\yv_K}$.

We note that this is a rather simplified model of a general decentralized multi-agent optimization problem. For the sake of exposition, we have introduced several implicit limitations:
\begin{enumerate}
\item We do not allow the agents to exchange information before/while taking their decisions. Each agent would otherwise optimize the information to share with the other agents, possibly in multiple rounds, jointly with the decision functions. In our illustrating example in Section~\ref{se:decentralized_ML:application}, we will do a slight abuse and allow for a $1$-stage cooperation link as it gives rise to interesting insights.
\item We have not taken into account the possibility to receive feedback from the environment after an action, e.g. in the form of an instantaneous reward. In that case, the agent would be able to gather information about the environment through its interactions and/or multiple agents might be able to coordinate their actions through this feedback. Such framework leads to the so-called \emph{reinforcement learning} approaches \cite{Sutton1998} that have been very successfully used in a large number of problems, in particular in combination with deep learning \cite{Mnih2015}. Yet, such approaches lead to other problems related to the convergence, the stability, and the efficiency of the equilibrium reached \cite{Sartoretti2019}. 
\end{enumerate}
Although written as a simple optimization problem, \eqref{eq:TD_1} models many scenarios where distributed intelligence needs to be efficiently used to reach the desired performance. It also entails a large complexity due to the specific functional dependencies which makes it necessary to optimize over multiple functional spaces of infinite dimensions.

We will now review conventional state-of-the-art approaches to this problem, and discuss why they do not exploit the decentralized intelligence. 
\paragraph*{Naive Strategy}
In the so-called \emph{naive} approach, each agent assumes that its information about the world is perfect, and all the other agents share the same information. Intuitively, this means that the decentralized configuration is not taken into account by the TX which runs a conventional centralized optimization algorithm.

Hence, the optimization problem solved by agent~$j$ is 
\begin{equation}
\left(\ldots,\bs_j^{\mathrm{naive}}\!\!\!\!\!\!,\ldots\right )\!=\!\argmax_{\bs_1,\ldots,\bs_n}  \mathrm{E}\left [ \!u(\xv,\bs_1\!(\yv_j),\ldots,\bs_n(\yv_j))\! \right ]\!.
\label{eq:TD_3}
\end{equation}   
This approach can be improved by taking into account the imperfection in $\yv_j$ with respect to $\xv$, i.e., taking the expectation over $p_{\xv\yv_j}$ as conventionally done in robust signal processing~\cite{Awan2018} instead of simply taking $\yv_j$ as being perfect, i.e., $\yv_j=\xv$. 
Yet, it is still fundamentally limited as the decentralized information structure is not taken into account: Coordination cannot be reached.
\paragraph*{Best Response}
A \emph{best-response} strategy is the optimal one given the strategies of the other agents, i.e., a  \emph{Nash equilibrium} \cite{Nash1951}. Hence, best-response strategies $(\bs^{\mathrm{BR}}_1,\ldots,\bs^{\mathrm{BR}}_n)$ satisfy:
\begin{equation}
\bs^{\mathrm{BR}}_j\!=\argmax_{\bs_j}\mathrm{E}\left[ u(\xv, \bs_j,\bar{\bs}^{\mathrm{BR}}_{j})  \!\right],
\label{eq:TD_4}
\end{equation} 
where we have used~$\bar{\bs}^{\mathrm{BR}}_{j}$ as a short-hand notations for all strategies~$\bs^{\mathrm{BR}}_{k}$ except $k=j$, and we have omitted the functional dependencies for the sake of clarity. A best-response strategy is also called a \emph{per-agent optimal strategy} and can be reached by finding the best strategy at each agent, and iterating over the agents. 

The iterations through the agents transform the decentralized optimization~\eqref{eq:TD_1} into a succession of conventional centralized functional optimization problems that can hence be tackled with conventional optimization tools. Yet, this best-response solution suffers from two important limitations. First, the best-response cooperation is a \emph{weak form of cooperation} as solutions necessitating a tight inter-dependency between the agents cannot be reached. More specifically, in a best-response approach, each agent can only update \emph{unilateraly}. This means that a solution necessitating several agents to update their strategies at the same time \emph{cannot} be reached. Second, it still requires solving a functional optimization problem at each agent, and is hence severely limited by the complexity when the dimension of the estimate increases.


\subsubsection{Centralized Training of Decentralized Strategies}
We will now discuss how the approach proposed in \cite{dekerret2018_ISWCS,Kim2018} allows to leverage recent progresses in ML, and in particular deep learning, to solve the two main challenges, which are (i) achieving a strong form of cooperation and  (ii) dealing with the complexity.

As a first step, optimizing \eqref{eq:TD_1} over the space of functions, it is natural to resort to a set of basis functions to reduce the dimensionality of the optimization space. Hence, we propose to restrict the strategy of agent~$j$ to belong to a parameterized subspace, i.e., to be of the form $\bs^{\bm{\theta}_j}_j$ with $\bm{\theta}_j$ being a vector of real parameters. We will  consider DNNs for their many advantages, and in particular, for the efficient implementations and the abundant literature \cite{Lecun2012,Goodfellow-et-al-2016}, but other ML methods could also be considered. The optimization problem~\eqref{eq:TD_1} is then approximated to:
\begin{equation}
\left ( \bm{\theta}_1^{\star},\ldots, \bm{\theta}_n^{\star}\right)\!=\!\argmax_{\bm{\theta}_1,\ldots,\bm{\theta}_n} \mathrm{E}\left [ u(\xv,\bs_1^{\bm{\theta}_1}(\yv_1),\ldots,\bs_n^{\bm{\theta}_n}(\yv_n)) \!\right]\!.
\label{eq:TD_2}
\end{equation}  
We then follow a data-driven approach, and aim at maximizing the average performance using samples from the joint distribution. This is possible as the objective utility function~$u$ is known and differentiable. This will be achieved by \emph{centralized training} to optimize over \emph{decision functions} using a (large amount) of available training samples. This means that, before evaluating how well our strategies perform in our decentralized environment by computing the objective in \eqref{eq:TD_1}, i.e., \emph{decentralized testing}, we consider an optimization phase where every information is available everywhere, i.e., \emph{centralized training}. 
\begin{remark}
\textbf{An analogy:} This centralized training/decentralized testing can be intuitively understood by considering football players within a team. During training, they can discuss and observe what the others are doing, they can see whether one gesture works or not, and test any move and strategy until it works. Yet, during the performance evaluation (i.e., the game) against a new opponent, they can use only what they observe to take the best action based on the tactics they have developed during training, and they can only exchange limited amount of information.\qed
\end{remark}

\begin{figure}[htp!] 
\centering
\includegraphics[width=1\columnwidth]{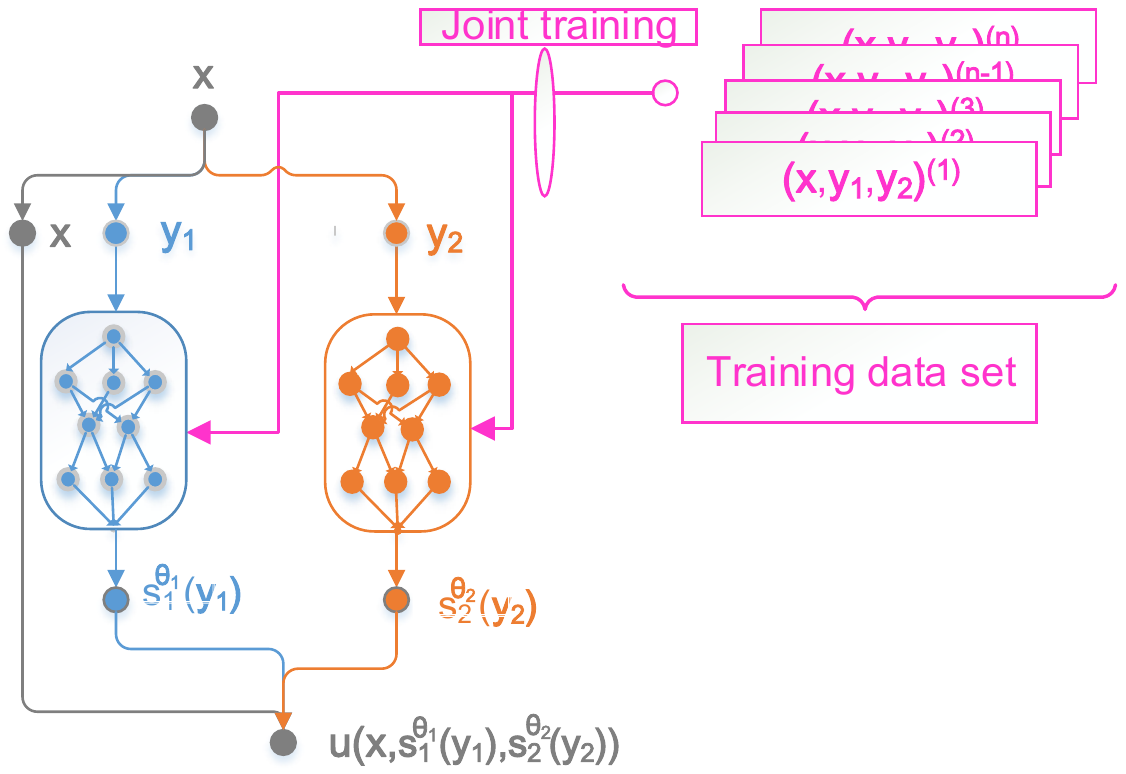}
\caption{Illustration of the decentralized use of DNNs with centralized training and decentralized testing.} 
\label{Tutorial_ML_TDNN}
\end{figure}

In practice, this means that we will jointly update the parameter vectors of all agents~$(\bm{\theta}_1,\ldots, \bm{\theta}_n)$ using the stochastic gradient approach during the training phase, as is standard in deep learning, i.e., at step~$k$
\begin{align}
&(\bm{\theta}_1^{(k)},\ldots, \bm{\theta}_n^{(k)})=(\bm{\theta}_1^{(k-1)},\ldots, \bm{\theta}_n^{(k-1)})+\notag\\
&~~~\alpha \nabla_{(\bm{\theta}_1,\ldots, \bm{\theta}_n)}u(\xv,\bs_1^{\bm{\theta}^{(k-1)}_1}(\yv_1),\ldots,\bs_n^{\bm{\theta}^{(k-1)}_n}(\yv_n))|.
\label{eq:TD_5} 
\end{align}
In \eqref{eq:TD_5}, we have written the gradient step at a single sample, i.e., the stochastic gradient. Any other more sophisticated gradient method (e.g., AdamGrad, Adagrad,...) can be used, and this easily done using the implementation packages available (e.g., Tensorflow). We illustrate the proposed decentralized use of DNNs in Fig.~\ref{Tutorial_ML_TDNN}.

As mentioned above, the advantage of the proposed formulation is that it makes it possible to apply many methods, algorithms, and implementations from the deep learning literature. Yet, it remains an open problem to determine how efficient these methods designed in a \emph{centralized setting} work, and whether they can be tailored to the decentralized problem at hand to improve the performance. In particular, it is an interesting open question to understand whether some architectures are better suited to the decentralized setting and how to improve the efficiency of training.

\subsubsection{Application in Wireless Networks: Learning to Cooperate in Coordinated Power Control}\label{se:decentralized_ML:application}
To illustrate the application of the concepts described above in wireless networks, we consider a toy example consisting of $2$ single-antenna transmitters (TXs) serving $2$ single-antenna receivers (RXs) in the same resource block with the goal of maximizing the sum rate. Specifically, let us denote the channel gain between TX~$i$ and RX~$j$ by $g_{i,j}$, the channel matrix containing all four channel coefficients by $\bG \in \mathbb{R}^{+}$, and the power transmitted by TX~$i$ by $p_i$. Considering unit-variance Gaussian noise, the instantaneous rate of user~$i$ can be written as 
\begin{equation} 
R_i(\bG,p_1,p_2)\! =\frac{1}{2}\log_2 \left(\! 1+\frac{g_{i,i}p_i}{1+g_{i,j}p_j} \right)\quad,j\neq i.
\end{equation}
We then define our sum rate utility function as $R=R_1+R_2$. Let us now assume that TX~$j$ has an imperfect estimate of that matrix, denoted by~$\mathbf{G}_j$ and that our optimization problem consists in finding the optimal power control strategies~$(p_1^{\star},p_2^{\star})$. Finally, We extend the previous formulation to allow \emph{a one stage} exchange of information between the two TXs, such that $p_1^{\star}$ is also a function of the information received from TX~$2$, which we denote by $z_{2,1}=f_{2,1}(\mathbf{G}_2)$ (and symmetrically). 

\begin{figure}[htp!] 
\centering
\includegraphics[width=1\columnwidth]{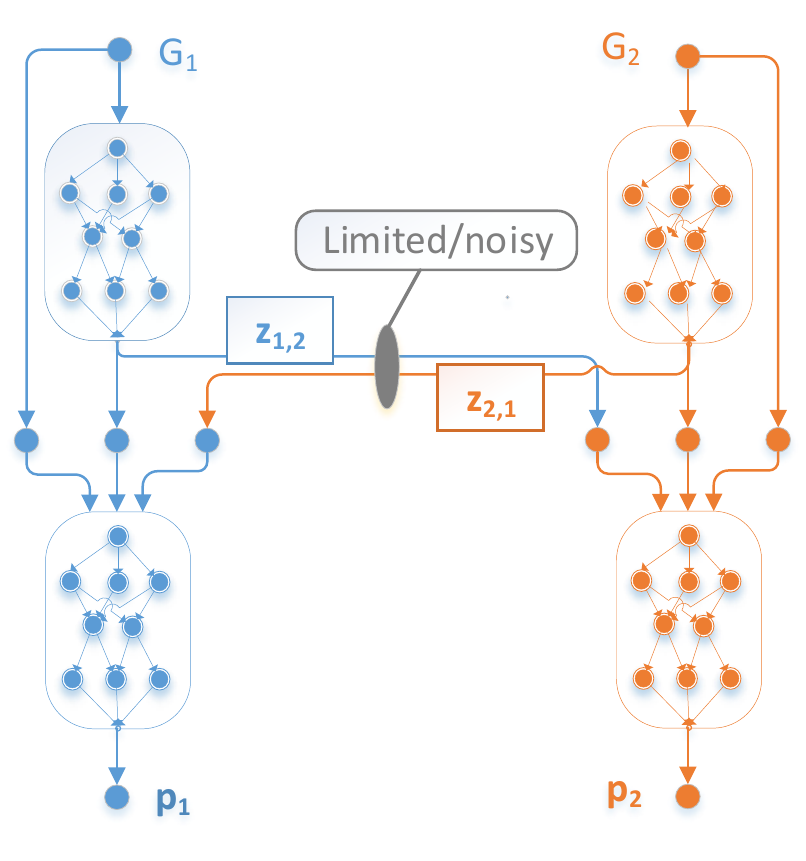}
\caption{Illustration of the decentralized DNN architecture with message exchange for power control.} 
\label{Tutorial_ML_SM}
\end{figure}

The optimization problem is then written as:
\begin{equation}  
(p_1^{\star}, p_2^{\star}) \!= \!\!\!\!\!\argmax\limits_{p_1, p_2,f_{1,2},f_{2,1}}\!\!\! \!\!\!
~\mathbb{E}\!\left[R\!\left(\mathbf{G},p_1(\mathbf{G}_1, \mathbf{z}_{2,1}),p_2(\mathbf{G}_2, \mathbf{z}_{1,2}))\right)\right].
\end{equation}
Following the data-driven approach described in the previous section, we use a different DNN to parameterize each decision function to be optimized, and we then train all DNNs jointly. The first two DNNs generate the messages to be shared for coordination, while the last two DNNs learn the power control from all inputs. This architecture is illustrated in Fig.~\ref{Tutorial_ML_SM}.

As discussed above, the training is done centrally following a conventional implementation of deep learning algorithms. The coefficients of all $4$~DNNs are updated together using the stochastic gradient or its variants. In the simulations, we  use a Tensorflow implementation of mini-batch gradient descent algorithm with the parameters given in Table~\ref{table_parameters}. To gain some insights into the algorithm, we consider a simple CSI configuration where TX~$1$ has a noisy CSIT, with all coefficients being corrupted by an independent Gaussian noise of variance~$\sigma^2$, and TX~$2$ has access to perfect CSI. To facilitate the qualitative interpretation, we furthermore reduce to an asymmetric setting where only TX~$1$ can share a message with TX~$2$ via the $1$-stage cooperation.

\begin{figure}[htp!] 
\centering
\includegraphics[width=1\columnwidth]{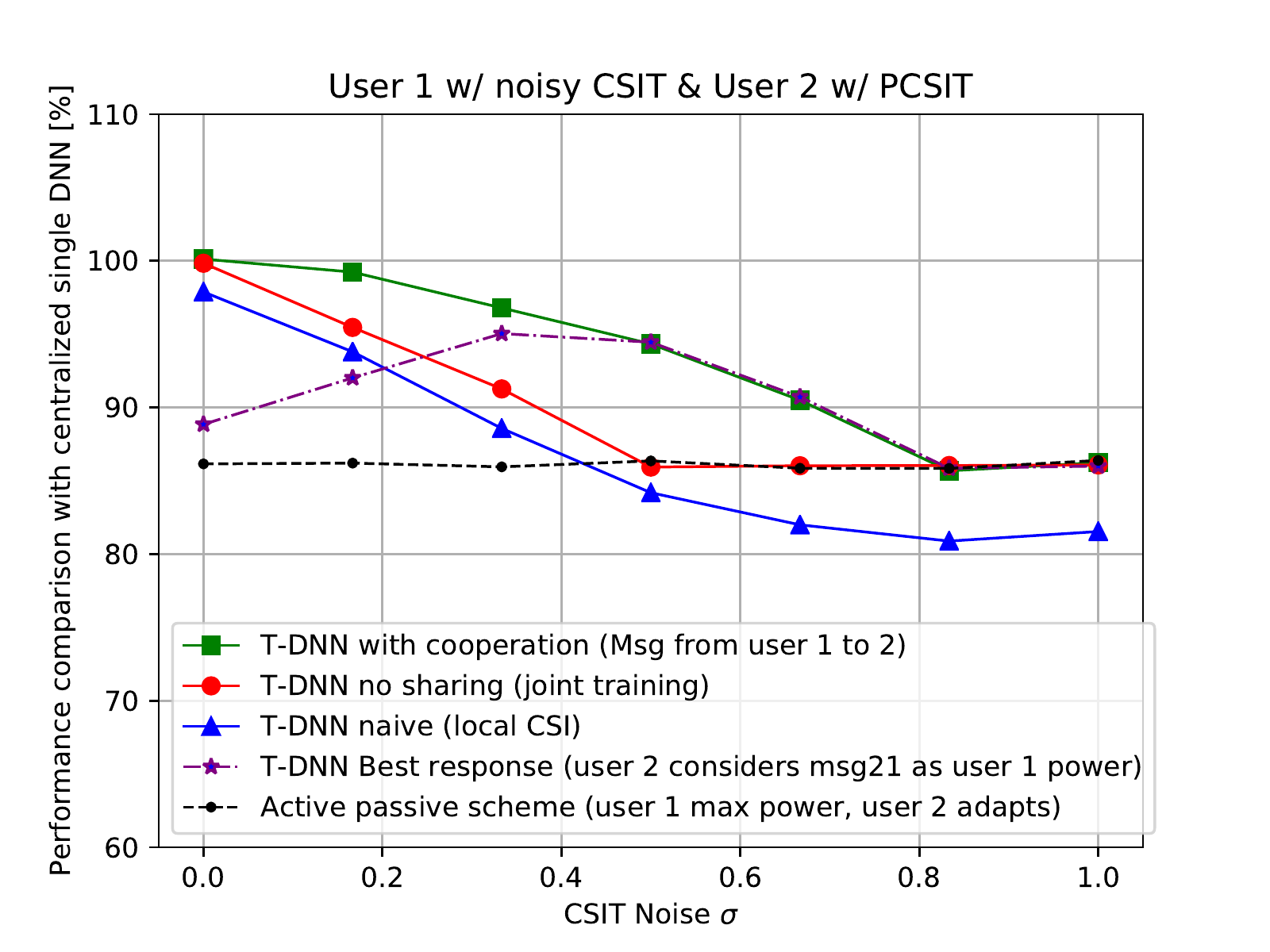}
\caption{Average sum rate as a function of the maximum transmit power~$P$.} 
\label{Tutorial_ML_rate}
\end{figure}

In Fig.~\ref{Tutorial_ML_rate}, we show the average sum rate achieved by different methods after normalization by the average sum rate achieved using deep learning in a centralized manner (with the same parameters), i.e., if perfect CSI is handed over to a central node controlling both TXs. This corresponds to the best that could be achieved using DNNs, and it has been shown in the literature \cite{Lee2018} that this approach outperforms other state-of-the-art schemes.

We present the performance after training of our decentralized DNN approach with and without a cooperation link. We then compare these two schemes to several reference schemes. First, we show the \emph{naive} use of DNNs in which each TX applies its learning algorithm as if he was the central controller of the two TX antennas (i.e., TX~$j$ uses only samples of $(\mathbf{G},\mathbf{G}_j)$). We can observe that our decentralized DNN approach outperforms the naive one, and that the algorithm learns to exploit the cooperation link available. 

We also compare with the so-called \emph{Active-Passive} scheme in which the less informed TX (TX~$1$) always transmits with full power, while the better-informed TX (TX~$2$) adapts in order to maximize the performance. When $\sigma=1$, i.e., when TX~$1$ has access to no CSI, the decentralized DNN approache converges to that scheme, as intuitively expected since TX~$1$ cannot exploit any instantaneous CSI knowledge.

Finally, we also consider a scheme in which TX~$1$ uses the trained decentralized DNN power control and message sharing strategy while TX~$2$ is forced to use the message received from TX~$1$ as being equal to the power used by TX~$1$. On the basis of this information, TX~$2$ then optimizes its power control. We can observe that this scheme achieves the same performance as the decentralized DNN scheme with cooperation when $\sigma\geq 0.5$. This hints towards the fact that TX~$1$ has learnt the strategy of sharing its power control decision to TX~$2$. How the cooperation link is exploited for lower values of~$\sigma$ is an interesting question under investigation.
\begin{table}[htp!]
\centering
\caption{Training parameters}
\begin{tabular}{c|c}
\hline
hidden layers& 4\\\hline
hidden nodes per layer& 50\\\hline
activation function& ReLU\\\hline
batch size& 400\\\hline
learning rate& 0.00003 (Adam)\\\hline
dropout rate in training& 0.3\\\hline
dropout rate in test& 0.0\\\hline
number of training samples& 4000\\\hline
number of test data& 10000\\\hline
training epochs& 1000\\\hline 
SNR sharing link &10 dB \\\hline
\end{tabular}
\label{table_parameters}
\end{table}

\section{Learning at the Wireless Edge}\label{sec:edge_learning}
All the applications of ML in wireless communications presented so far are based on centralized training even if the implementation is distributed as explained in the previous section. However, as mentioned in Section \ref{s:intro}, centralized training can be challenging in certain wireless scenarios with limited bandwidth and energy resources, as this would require transmitting all the data to a cloud processor for centralized training, and often a single device is limited in terms of both available data and computation power. Privacy is another concern that can prevent centralized ML for most sensor data collected by IoT devices, e.g., smart meters \cite{Giaconi:SM:18} or electric vehicles \cite{V2X:privacy}.  An alternative is to implement learning at the wireless edge, also called \textit{edge learning} \cite{edge:bennis}, in the form of distributed stochastic gradient descent (DSGD) or federated learning (FL) \cite{FL:KonecnyMRR16}.

It is now commonly accepted that the main bottleneck in distributed learning is the communication load \cite{DL_CommsEff}. Due to the lack of centralized processing capabilities, these algorithms depend heavily on information exchange between multiple learning agents, called \textit{workers}, either through a `master' orchestrating node, called a \textit{parameter server}, or in a fully distributed fashion through device-to-device communications. In either case, distributed learning requires iterative information exchange among the workers and the parameter server, where the workers share either their local gradient estimates in DSGD, or their local model estimates in FL. 

There have been numerous studies that focus on communication-efficient distributed ML. These studies can be grouped into three different approaches, namely \textit{quantization}, \textit{sparsification}, and \textit{local updates}. Quantization algorithms aim at reducing the amount of information that need to be communicated to convey the result of local learning iteration, e.g., the local gradient estimate \cite{DCLimitedPrecisionGupta,DCOneBitQuan}. Sparsification, on the other hand, reduces the communication load by transmitting only the important values of local estimates \cite{ScalableDNNStorm,DCAjiSparse,DCSattlerSparseBinary}. Another approach is to reduce the frequency of
communication from the workers by allowing local parameter
updates \cite{McMahan2017CommunicationEfficientLO,LAGGradientGiannakis}. We remark, however, that, these studies do not explicitly model the underlying communication channel between the workers and the parameter server, and mainly focus on large scale distributed learning within server farms, where hundreds, maybe thousands of machines collaborate to learn a high-dimensional model on an extremely large dataset. However, as we will show below, taking the particular channel model into account is critical in wireless edge learning, where the channel can be severely limiting. 

\begin{figure} 
\centering
\includegraphics[width=1\columnwidth]{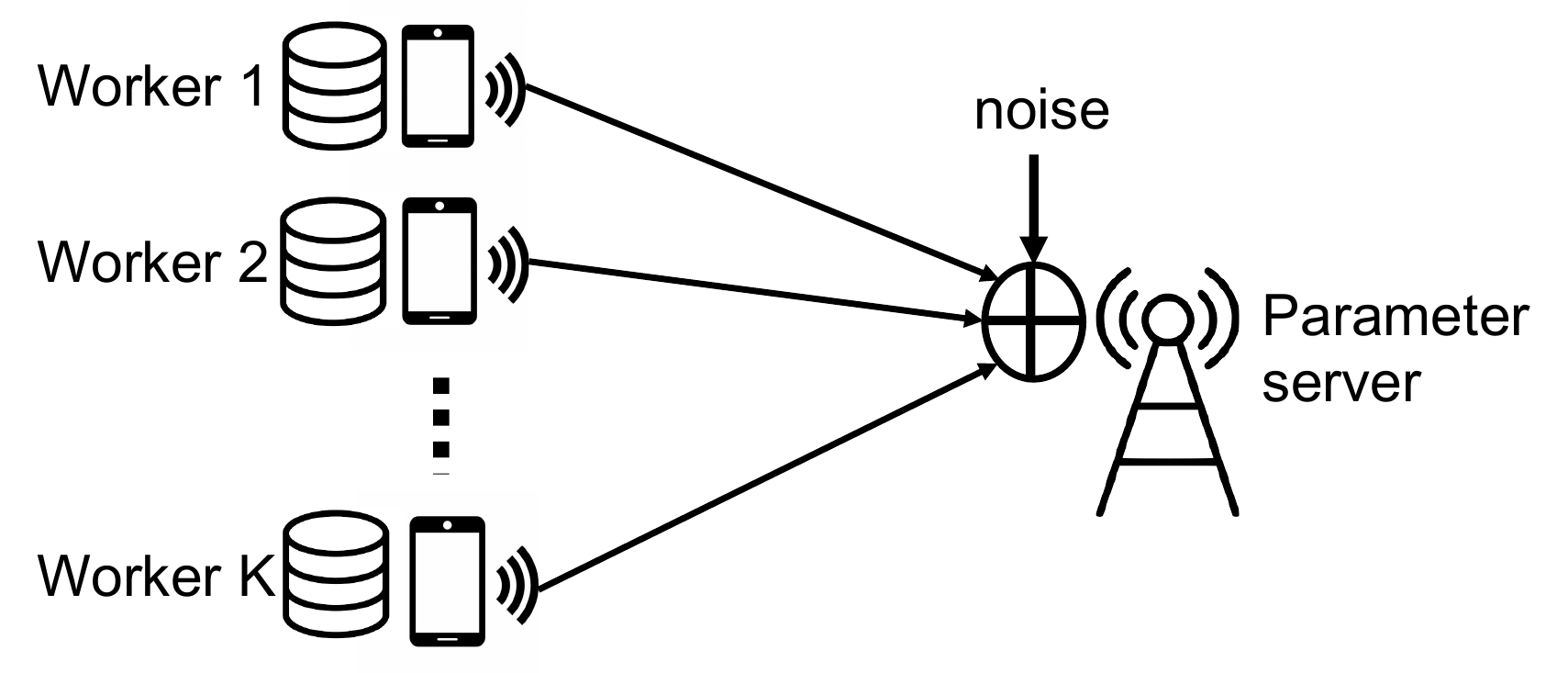}
\caption{Wireless edge learning where the workers communicate their local estimates to the parameter server over a shared wireless channel.} 
\label{fig:edge_learning}
\end{figure}

Consider DSGD over a shared wireless medium, as illustrated in Fig. \ref{fig:edge_learning}, where the transmission of local gradient estimates from the workers to the parameter server can be formulated as a wireless computation problem \cite{Amiri:edge}. One approach to this problem is to treat communication and computation separately, and exploit a coding scheme across computing agents such that each of them is assigned a non-zero rate to convey its gradient estimate at each iteration. Therefore, each agent employs quantization to reduce the amount of information to be transmitted to the level that is allowed by the wireless channel. This can be called a `separate digital' scheme as the gradient estimates are converted into bits, which are communicated by independent channel codes. 

Note, however, that, the parameter server is interested only in the average of the gradient estimates, rather than their individual values. Accordingly, a much more efficient communication strategy would be to transmit local estimates without any coding, in an `analog' fashion. If all the workers are synchronized, than the wireless channel adds their estimates, directly conveying the desired value to the parameter server (which simply divides this sum by the number of workers to find the average). A random projection of the gradient estimates is proposed in \cite{Amiri:edge} to reduce the required channel bandwidth. This approach can also be extended to the scenario with fading \cite{Zhu:Huang:edge, Amiri:fading:edge}, in which case power control can be employed at the workers to align their transmissions at the same received power level.  

\begin{figure} 
\centering
\includegraphics[width=1\columnwidth]{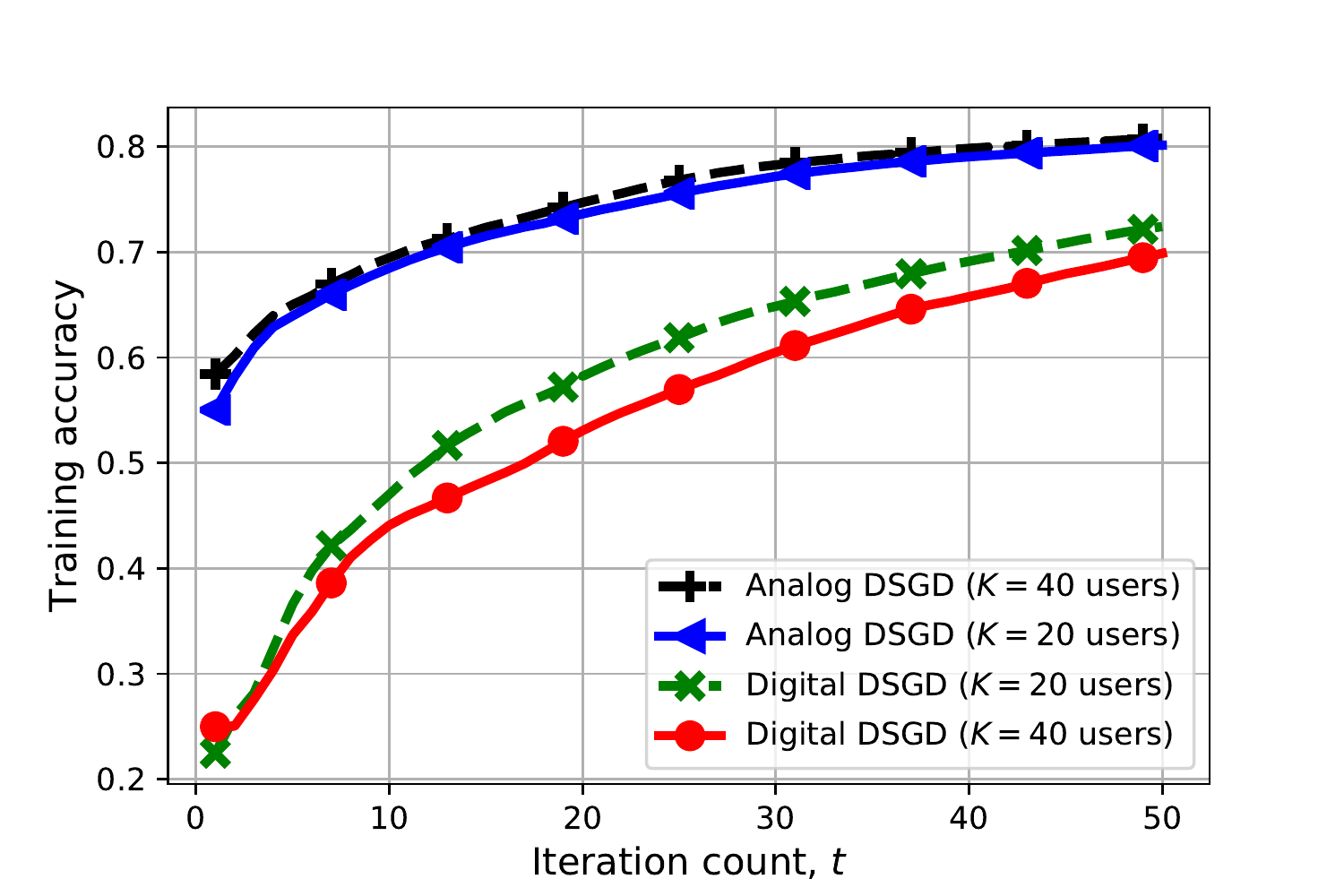}
\caption{Test accuracy of analog and digital transmission schemes for wireless edge learning with different number of users.} 
\label{fig:analog_vs_digital}
\end{figure}

In Fig. \ref{fig:analog_vs_digital} we illustrate the performance of the digital and analog computation approaches for learning over the wireless edge. The figure compares the test accuracy when a
single layer NN is trained on the MNIST dataset. A total of $60000$ data samples are distributed across $K$ workers, which employ DSGD utilizing ADAM optimizer. The figure compares the accuracy achieved for a fixed average transmit power value for each user. We observe that analog transmission of gradient estimates achieves a significantly higher accuracy compared to first quantizing the estimates, and then transmitting the quantized bits with a channel code. We also make an interesting observation from Fig.~\ref{fig:analog_vs_digital}: while the accuracy of the analog scheme increases with the number of workers, as each additional worker comes with its own power source, the digital scheme has an optimal number of workers, beyond which the accuracy degrades. This is because, channel resources per worker becomes limited beyond this optimal number of workers, which, in turn, limits the accuracy of the gradient estimates that are conveyed to the parameter server. 

Overall, the results highlight the fact that, for efficient ML at the wireless edge, communication and computation have to be considered jointly, and distributed ML can benefit from physical layer techniques to improve the efficiency and accuracy. A similar observation is also made in \cite{simeone:wirelessComputation} by considering coded wireless computation in the map-shuffle-reduce framework, where physical layer techniques are leveraged to provide robustness against worker and channel uncertainties.

\section{Conclusions}\label{se:decentralized_ML:discussion}


We have presented what we hope is a stimulating overview of the promises and challenges of ML for the physical layer of wireless networks. We have presented a wide variety of wireless communications problems, in which ML tools have been shown to offer significant gains. As indicated earlier, these correspond to scenarios in which either we do not have an accurate model of the system, or we have an accurate model but the optimal solution is extremely complex and thus cannot be attained with conventional means. Various power allocation problems have been presented as good examples of the latter scenario. The joint source-channel coding problem can be considered as exhibiting both limitations. In the case of image transmission, we do not have a good statistical model of natural images; however, even when transmitting Gaussian sources over a noisy channel, the optimal solution is not known for finite blocklengths, as the separation theorem fails. We have also highlighted another connection between ML and the wireless physical layer through edge learning. We have shown that the accuracy of distributed ML over wireless channels can benefit greatly from the joint treatment of the physical layer and the employed learning algorithm.



In light of these intriguing achievements and challenges, an important question that our research community will tackle in the next few years is the following: Will the ongoing ML revolution completely transform communication system design, so that we will soon be designing \textit{autonomous communication devices} that do not need standards or protocols, and can simply learn to communicate with one another using data-driven ML techniques? Or do the existing drawbacks of ML-based techniques limit their relevance for communication systems, and we should instead ``stick to our guns'' -- the time-tested highly-optimized model-based approaches? While time will tell what the answer is, it will probably land somewhere in the middle; strong domain knowledge and model-based approaches will need to be combined with powerful data-driven ML techniques. Another important question relates to the use of physical layer techniques for edge learning: Given the rapid speed of developments in ML and particularly edge learning, do we need new standards and new communication techniques that can sustain the growing demand for ML applications at the edge?

We believe that ML and data-driven approaches in general have a lot to offer to all aspects of the communication network architecture, and they have already started to have impact on the higher layers \cite{Li:AI_5G:17, Kibria:BigData:5G, Kafle:Slicing:18}. Yet, to realize this promise, significant research efforts are needed, from adaptation of existing ML techniques to the development of new ones that can meet the constraints and requirements of communication networks, including the implementation of at least some of these capabilities in low-power chips that can be used in mobile devices \cite{Venieris:DNN_Embedded:18, Yoo:AI_Silicon:19}, and/ or the development of fully distributed, yet efficient  implementations that can employ low-power low-complexity mobile devices. 

To conclude, one message that comes out loud and clear from our recent experience with deep learning and other data-driven approaches is that we should think big and be bold. Communications engineers are trained to think about physical models and optimal solutions, but the success of deep learning hinges on using lots of data together with `naive' lightweight approaches, like stochastic gradient descent, to solve NP-hard problems. It takes quite a bit of cultural transformation to digest this. Moreover, the performance of these generic lightweight tools can be improved significantly through domain expertise in wireless communications, complemented with a thorough knowledge of the tricks of the trade in ML.

 
\bibliographystyle{IEEEtran}
\bibliography{Paper}

\begin{thebibliography}{10}
\providecommand{\url}[1]{#1}
\csname url@samestyle\endcsname
\providecommand{\newblock}{\relax}
\providecommand{\bibinfo}[2]{#2}
\providecommand{\BIBentrySTDinterwordspacing}{\spaceskip=0pt\relax}
\providecommand{\BIBentryALTinterwordstretchfactor}{4}
\providecommand{\BIBentryALTinterwordspacing}{\spaceskip=\fontdimen2\font plus
\BIBentryALTinterwordstretchfactor\fontdimen3\font minus
  \fontdimen4\font\relax}
\providecommand{\BIBforeignlanguage}[2]{{%
\expandafter\ifx\csname l@#1\endcsname\relax
\typeout{** WARNING: IEEEtran.bst: No hyphenation pattern has been}%
\typeout{** loaded for the language `#1'. Using the pattern for}%
\typeout{** the default language instead.}%
\else
\language=\csname l@#1\endcsname
\fi
#2}}
\providecommand{\BIBdecl}{\relax}
\BIBdecl

\bibitem{Goodfellow-et-al-2016}
I.~Goodfellow, Y.~Bengio, and A.~Courville, \emph{Deep Learning}.\hskip 1em
  plus 0.5em minus 0.4em\relax MIT Press, 2016,
  \url{http://www.deeplearningbook.org}.

\bibitem{Sutton2018}
R.~S. Sutton and A.~G. Barto, \emph{{Reinforcement Learning: An
  Introduction}}.\hskip 1em plus 0.5em minus 0.4em\relax Second Edition, MIT
  Press, Cambridge, MA, 2018.

\bibitem{Shannon}
C.~E. Shannon, ``A mathematical theory of communication,'' \emph{Bell Syst.
  Technical Journal}, vol.~27, no.~3, pp. 379--423, Jul 1948.

\bibitem{Gersho:vecquant}
A.~Gersho and R.~M. Gray, \emph{Vector Quantization and Signal
  Compression}.\hskip 1em plus 0.5em minus 0.4em\relax Norwell, MA, USA: Kluwer
  Academic Publishers, 1991.

\bibitem{LZ77}
J.~{Ziv} and A.~{Lempel}, ``A universal algorithm for sequential data
  compression,'' \emph{IEEE Transactions on Information Theory}, vol.~23,
  no.~3, pp. 337--343, May 1977.

\bibitem{Willems:CWT}
F.~M.~J. {Willems}, Y.~M. {Shtarkov}, and T.~J. {Tjalkens}, ``The context-tree
  weighting method: basic properties,'' \emph{IEEE Transactions on Information
  Theory}, vol.~41, no.~3, pp. 653--664, May 1995.

\bibitem{Hamed:survey}
\BIBentryALTinterwordspacing
C.~Zhang, P.~Patras, and H.~Haddadi, ``Deep learning in mobile and wireless
  networking: {A} survey,'' \emph{CoRR}, vol. abs/1803.04311, 2018. [Online].
  Available: \url{http://arxiv.org/abs/1803.04311}
\BIBentrySTDinterwordspacing

\bibitem{NN_Tutorial_CHen}
\BIBentryALTinterwordspacing
M.~Chen, U.~Challita, W.~Saad, C.~Yin, and M.~Debbah, ``Machine learning for
  wireless networks with artificial intelligence: {A} tutorial on neural
  networks,'' \emph{CoRR}, vol. abs/1710.02913, 2017. [Online]. Available:
  \url{http://arxiv.org/abs/1710.02913}
\BIBentrySTDinterwordspacing

\bibitem{ML_for_VN}
L.~{Liang}, H.~{Ye}, and G.~Y. {Li}, ``Toward intelligent vehicular networks: A
  machine learning framework,'' \emph{IEEE Internet of Things Journal}, vol.~6,
  no.~1, pp. 124--135, Feb 2019.

\bibitem{Jagannath:Survey:IoT}
\BIBentryALTinterwordspacing
J.~Jagannath, N.~Polosky, A.~Jagannath, F.~Restuccia, and T.~Melodia, ``Machine
  learning for wireless communications in the internet of things: {A}
  comprehensive survey,'' \emph{CoRR}, vol. abs/1901.07947, 2019. [Online].
  Available: \url{http://arxiv.org/abs/1901.07947}
\BIBentrySTDinterwordspacing

\bibitem{Zhou_CR_ML}
X.~{Zhou}, M.~{Sun}, G.~Y. {Li}, and B.~{Fred Juang}, ``Intelligent wireless
  communications enabled by cognitive radio and machine learning,'' \emph{China
  Communications}, vol.~15, no.~12, pp. 16--48, Dec 2018.

\bibitem{Simeone_TCCN_2018}
O.~Simeone, ``{A very brief introduction to Machine Learning with applications
  to communication systems},'' \emph{IEEE Trans. on Cognitive Commun. and
  Networking}, vol.~4, no.~4, pp. 648--664, Dec. 2018.

\bibitem{interpret:arxiv:17}
F.~Doshi-Velez and B.~Kim, ``{Towards A Rigorous Science of Interpretable
  Machine Learning},'' \emph{arXiv e-prints}, p. arXiv:1702.08608, Feb. 2017.

\bibitem{OShea:GNU:16}
\BIBentryALTinterwordspacing
T.~O'Shea and N.~West, ``Radio machine learning dataset generation with gnu
  radio,'' \emph{Proceedings of the GNU Radio Conference}, vol.~1, no.~1, 2016.
  [Online]. Available:
  \url{https://pubs.gnuradio.org/index.php/grcon/article/view/11}
\BIBentrySTDinterwordspacing

\bibitem{Alkhateeb2019}
A.~Alkhateeb, ``Deep{MIMO}: A generic deep learning dataset for millimeter wave
  and massive {MIMO} applications,'' in \emph{Proc. of Information Theory and
  Applications Workshop (ITA)}, San Diego, CA, Feb 2019, pp. 1--8.

\bibitem{Brazil:dataset}
I.~Nascimento, F.~Mendes, M.~Dias, A.~Silva, and A.~Klautau, ``Deep learning in
  rat and modulation classification with a new radio signals dataset,'' in
  \emph{Proc. XXXVI Simposio Brasileiro de Telecomunicacoes e Processamento de
  Sinais (SBrT)}, Brazil, Sep. 2018.

\bibitem{Arnold:dataset}
M.~Arnold, J.~Hoydis, and S.~ten Brink, ``Novel massive {MIMO} channel sounding
  data applied to deep learning-based indoor positioning,'' in \emph{Proc.
  Int'l ITG Conf. on Systems, Communications and Coding (SCC)}, Feb. 2019.

\bibitem{LearnWMMSE}
H.~Sun, X.~Chen, Q.~Shi, M.~Hong, X.~Fu, and N.~D. Sidiropoulos, ``Learning to
  optimize: Training deep neural networks for interference management,''
  \emph{IEEE Transactions on Signal Processing}, vol.~66, no.~20, pp.
  5438--5453, Oct 2018.

\bibitem{Vinyals:NIPS:15}
\BIBentryALTinterwordspacing
O.~Vinyals, M.~Fortunato, and N.~Jaitly, ``Pointer networks,'' in
  \emph{Advances in Neural Information Processing Systems 28}, C.~Cortes, N.~D.
  Lawrence, D.~D. Lee, M.~Sugiyama, and R.~Garnett, Eds.\hskip 1em plus 0.5em
  minus 0.4em\relax Curran Associates, Inc., 2015, pp. 2692--2700. [Online].
  Available: \url{http://papers.nips.cc/paper/5866-pointer-networks.pdf}
\BIBentrySTDinterwordspacing

\bibitem{NPhard:app}
\BIBentryALTinterwordspacing
A.~Milan, S.~Rezatofighi, R.~Garg, A.~Dick, and I.~Reid, ``Data-driven
  approximations to np-hard problems,'' 2017. [Online]. Available:
  \url{https://aaai.org/ocs/index.php/AAAI/AAAI17/paper/view/14700}
\BIBentrySTDinterwordspacing

\bibitem{Atari:DL}
\BIBentryALTinterwordspacing
V.~Mnih, K.~Kavukcuoglu, D.~Silver, A.~Graves, I.~Antonoglou, D.~Wierstra, and
  M.~A. Riedmiller, ``Playing atari with deep reinforcement learning,''
  \emph{CoRR}, vol. abs/1312.5602, 2013. [Online]. Available:
  \url{http://arxiv.org/abs/1312.5602}
\BIBentrySTDinterwordspacing

\bibitem{Silver:AGoZ}
\BIBentryALTinterwordspacing
D.~Silver, J.~Schrittwieser, K.~Simonyan, I.~Antonoglou, A.~Huang, A.~Guez,
  T.~Hubert, L.~Baker, M.~Lai, A.~Bolton, Y.~Chen, T.~Lillicrap, F.~Hui,
  L.~Sifre, G.~van~den Driessche, T.~Graepel, and D.~Hassabis, ``Mastering the
  game of go without human knowledge,'' \emph{Nature}, vol. 550, pp. 354 EP --,
  10 2017. [Online]. Available: \url{https://doi.org/10.1038/nature24270}
\BIBentrySTDinterwordspacing

\bibitem{Farsad_TSP_2018}
N.~{Farsad} and A.~{Goldsmith}, ``{Neural network detection of data sequences
  in communication systems},'' \emph{IEEE Trans. Signal Process.}, vol.~66,
  no.~21, pp. 5663--5678, Nov. 2018.

\bibitem{Samuel2017}
\BIBentryALTinterwordspacing
N.~Samuel, T.~Diskin, and A.~Wiesel, ``{Deep MIMO detection},'' 2017. [Online].
  Available: \url{https://arxiv.org/pdf/1706.01151.pdf}
\BIBentrySTDinterwordspacing

\bibitem{Ye_WComLet_2018}
H.~Ye, G.~Y. Li, and B.~Juang, ``{Power of deep learning for channel estimation
  and signal detection in OFDM systems},'' \emph{IEEE Wireless Comm. Letters},
  vol.~7, no.~1, pp. 114--117, Feb. 2018.

\bibitem{Jiang_ArXiv_2018}
P.~{Jiang}, T.~{Wang}, B.~{Han}, X.~{Gao}, J.~{Zhang}, C.-K. {Wen}, S.~{Jin},
  and G.~Y. {Li}, ``{Artificial Intelligence-aided OFDM Receiver: Design and
  Experimental Results},'' \emph{arXiv e-prints}, p. arXiv:1812.06638, Dec.
  2018.

\bibitem{Neumann2018LearningTM}
D.~Neumann, T.~Wiese, and W.~Utschick, ``Learning the mmse channel estimator,''
  \emph{IEEE Transactions on Signal Processing}, vol.~66, pp. 2905--2917, 2018.

\bibitem{Soltani_ArXiv_2019}
\BIBentryALTinterwordspacing
M.~Soltani, A.~Mirzaei, V.~Pourahmadi, and H.~Sheikhzadeh, ``Deep
  learning-based channel estimation,'' \emph{CoRR}, vol. abs/1810.05893v2,
  2010. [Online]. Available: \url{http://arxiv.org/abs/1810.05893}
\BIBentrySTDinterwordspacing

\bibitem{Koller2018MachineLF}
M.~Koller, C.~Hellings, M.~Knoedlseder, T.~Wiese, D.~Neumann, and W.~Utschick,
  ``Machine learning for channel estimation from compressed measurements,''
  \emph{2018 15th International Symposium on Wireless Communication Systems
  (ISWCS)}, pp. 1--5, 2018.

\bibitem{He_ArXiv_2018}
\BIBentryALTinterwordspacing
H.~He, C.~Wen, S.~Jin, and G.~Y. Li, ``Deep learning-based channel estimation
  for beamspace mmwave massive {MIMO} systems,'' \emph{CoRR}, vol.
  abs/1802.01290, 2018. [Online]. Available:
  \url{http://arxiv.org/abs/1802.01290}
\BIBentrySTDinterwordspacing

\bibitem{Nachmani2016}
E.~Nachmani, Y.~Be'ery, and D.~Burshtein, ``{Learning to decode linear codes
  using deep learning},'' in \emph{Proc. {A}llerton {C}onference on
  {C}ommunication, {C}ontrol, and {C}omputing (Allerton)}, 2016.

\bibitem{Cammerer2017}
\BIBentryALTinterwordspacing
S.~Cammerer, T.~Gruber, J.~Hoydis, and S.~ten Brink, ``{Scaling deep
  learning-based decoding of polar codes via partitioning},'' 2017. [Online].
  Available: \url{https://arxiv.org/abs/1702.06901}
\BIBentrySTDinterwordspacing

\bibitem{Gruber2017}
T.~Gruber, S.~Cammerer, J.~Hoydis, and S.~ten Brink, ``{On deep learning-based
  channel decoding},'' in \emph{Proc. Conference on Information Sciences and
  Systems (CISS)}, 2017.

\bibitem{Felix_ArXiv_2018}
\BIBentryALTinterwordspacing
A.~Felix, S.~Cammerer, S.~D{\"{o}}rner, J.~Hoydis, and S.~ten Brink,
  ``{OFDM}-autoencoder for end-to-end learning of communications systems,''
  \emph{CoRR}, vol. abs/1803.05815, 2018. [Online]. Available:
  \url{http://arxiv.org/abs/1803.05815}
\BIBentrySTDinterwordspacing

\bibitem{Raj_ComLet_2018}
V.~Raj and S.~Kalyani, ``Backpropagating through the air: Deep learning at
  physical layer without channel models,'' \emph{IEEE Communication Letters},
  vol.~22, no.~11, pp. 2278--2281, Nov. 2018.

\bibitem{Setiono:autoencoder}
R.~Setiono and G.~Lu, ``Image compression using a feedforward neural network,''
  in \emph{Proceedings of 1994 IEEE International Conference on Neural Networks
  (ICNN'94)}, vol.~7, June 1994, pp. 4761--4765 vol.7.

\bibitem{Balle:ICLR:17}
J.~Balle, V.~Laparra, and E.~P. Simoncelli, ``End-to-end optimized image
  compression,'' in \emph{Proc. of Int. Conf. on Learning Representations
  (ICLR)}, Apr. 2017, pp. 1--27.

\bibitem{Han:deepVideo}
J.~Han, S.~Lombardo, C.~Schroers, and S.~Mandt, ``Deep probabilistic video
  compression,'' \emph{CoRR}, vol. abs/1810.02845, 2018.

\bibitem{Oshea:16:arXiv}
\BIBentryALTinterwordspacing
T.~J. O'Shea, K.~Karra, and T.~C. Clancy, ``Learning to communicate: Channel
  auto-encoders, domain specific regularizers, and attention,'' \emph{CoRR},
  vol. abs/1608.06409, 2016. [Online]. Available:
  \url{http://arxiv.org/abs/1608.06409}
\BIBentrySTDinterwordspacing

\bibitem{Oshea2017}
T.~O'Shea and J.~Hoydis, ``{An introduction to deep learning for the physical
  layer},'' \emph{IEEE Trans. on Cognitive Communications and Networking},
  vol.~PP, no.~99, 2017.

\bibitem{Ye_ArXiv_2018}
\BIBentryALTinterwordspacing
H.~Ye, G.~Y. Li, B.~F. Juang, and K.~Sivanesan, ``Channel agnostic end-to-end
  learning based communication systems with conditional {GAN},'' \emph{CoRR},
  vol. abs/1807.00447, 2018. [Online]. Available:
  \url{http://arxiv.org/abs/1807.00447}
\BIBentrySTDinterwordspacing

\bibitem{Dorner2017}
\BIBentryALTinterwordspacing
S.~DÃ¶rner, S.~Cammerer, J.~Hoydis, and S.~ten Brink, ``{Deep learning-based
  communication over the air},'' 2017. [Online]. Available:
  \url{https://arxiv.org/abs/1707.03384}
\BIBentrySTDinterwordspacing

\bibitem{Eirina:deepJSCC}
\BIBentryALTinterwordspacing
E.~Bourtsoulatze, D.~B. Kurka, and D.~G{\"{u}}nd{\"{u}}z, ``Deep joint
  source-channel coding for wireless image transmission,'' \emph{CoRR}, vol.
  abs/1809.01733, 2018. [Online]. Available:
  \url{http://arxiv.org/abs/1809.01733}
\BIBentrySTDinterwordspacing

\bibitem{Cover2006}
T.~Cover and A.~Thomas, \emph{{Elements of information theory}}.\hskip 1em plus
  0.5em minus 0.4em\relax Wiley-Interscience, Jul. 2006.

\bibitem{Kurka:arxiv:19}
\BIBentryALTinterwordspacing
D.~B. Kurka and D.~G{\"{u}}nd{\"{u}}z, ``Successive refinement of images with
  deep joint source-channel coding,'' \emph{CoRR}, vol. abs/1903.06333, 2019.
  [Online]. Available: \url{http://arxiv.org/abs/1903.06333}
\BIBentrySTDinterwordspacing

\bibitem{LearnAntSel}
M.~S. Ibrahim, A.~S. Zamzam, X.~Fu, and N.~D. Sidiropoulos, ``Learning-based
  antenna selection for multicasting,'' in \emph{2018 IEEE 19th International
  Workshop on Signal Processing Advances in Wireless Communications (SPAWC)},
  June 2018, pp. 1--5.

\bibitem{LearnDSSE}
A.~S. {Zamzam}, X.~{Fu}, and N.~D. {Sidiropoulos}, ``{Data-Driven
  Learning-Based Optimization for Distribution System State Estimation},''
  \emph{arXiv e-prints}, p. arXiv:1807.01671, Jul. 2018.

\bibitem{BMF4MinOut}
V.~Ntranos, N.~D. Sidiropoulos, and L.~Tassiulas, ``On multicast beamforming
  for minimum outage,'' \emph{IEEE Transactions on Wireless Communications},
  vol.~8, no.~6, pp. 3172--3181, June 2009.

\bibitem{Learn2BMF}
Y.~Shi, A.~Konar, N.~D. Sidiropoulos, X.~Mao, and Y.~Liu, ``Learning to
  beamform for minimum outage,'' \emph{IEEE Transactions on Signal Processing},
  vol.~66, no.~19, pp. 5180--5193, Oct 2018.

\bibitem{Christensen2008}
S.~S. Christensen, R.~Agarwal, E.~Carvalho, and J.~M. Cioffi, ``{Weighted
  sum-rate maximization using weighted MMSE for MIMO-BC beamforming design},''
  \emph{IEEE Trans. on Wireless Commun.}, vol.~7, no.~12, pp. 4792--4799, Dec.
  2008.

\bibitem{Shi2011}
Q.~Shi, M.~Razaviyayn, Z.~Luo, and C.~He, ``{An iteratively weighted MMSE
  approach to distributed sum-utility maximization for a MIMO interfering
  Broadcast Channel},'' \emph{IEEE Trans. Signal Process.}, vol.~59, no.~9, pp.
  4331--4340, Sept 2011.

\bibitem{Kaleva2015}
J.~{Kaleva}, A.~{Tölli}, and M.~{Juntti}, ``{Successive convex approximation
  for simultaneous linear TX/RX design in MIMO BC},'' in \emph{Proc. IEEE
  Asilomar Conference on Signals, Systems and Computers (ACSSC)}, Nov. 2015.

\bibitem{LearnWMMSE-SPAWC}
H.~Sun, X.~Chen, Q.~Shi, M.~Hong, X.~Fu, and N.~D. Sidiropoulos, ``Learning to
  optimize: Training deep neural networks for wireless resource management,''
  in \emph{2017 IEEE 18th International Workshop on Signal Processing Advances
  in Wireless Communications (SPAWC)}, July 2017, pp. 1--6.

\bibitem{Lee2018}
W.~{Lee}, M.~{Kim}, and D.~{Cho}, ``{Deep power control: Transmit power control
  scheme based on convolutional neural network},'' \emph{IEEE Communications
  Letters}, vol.~22, no.~6, pp. 1276--1279, Jun. 2018.

\bibitem{Radner1962}
R.~Radner, ``{Team decision problems},'' \emph{The Annals of Mathematical
  Statistics}, 1962.

\bibitem{Marschak1972}
J.~Marschak and R.~Radner, \emph{{Economic theory of teams}}.\hskip 1em plus
  0.5em minus 0.4em\relax Yale University Press, New Haven and London, Feb.
  1972.

\bibitem{Sutton1998}
R.~S. Sutton and A.~G. Barto, \emph{{Reinforcement Learning: An
  Introduction}}.\hskip 1em plus 0.5em minus 0.4em\relax MIT Press, Cambridge,
  MA, 1998.

\bibitem{Mnih2015}
V.~Mnih, K.~Kavukcuoglu, D.~Silver, A.~A. Rusu, J.~Veness, M.~G. Bellemare,
  A.~Graves, M.~Riedmiller, A.~K. Fidjeland, G.~Ostrovski, S.~Petersen,
  C.~Beattie, A.~Sadik, I.~Antonoglou, H.~King, D.~Kumaran, D.~Wierstra,
  S.~Legg, and D.~Hassabis, ``{Human-level control through deep reinforcement
  learning},'' \emph{Nature International Weekly Journal of Science}, vol. 518,
  pp. 529--533, Feb. 2015.

\bibitem{Sartoretti2019}
G.~Sartoretti, Y.~Wu, W.~Paivine, T.~K.~S. Kumar, S.~Koenig, and H.~Choset,
  ``{Distributed reinforcement learning for multi-robot decentralized
  collective construction},'' \emph{Correll N., Schwager M., Otte M. (eds)
  Distributed Autonomous Robotic Systems. Springer Proceedings in Advanced
  Robotics}, vol.~9, 2019.

\bibitem{Awan2018}
D.~A. {Awan}, R.~L.~G. {Cavalcante}, and S.~{Stanczak}, ``{A robust machine
  learning method for cell-load approximation in wireless networks},'' in
  \emph{Proc. IEEE International Conference on Acoustics, Speech and Signal
  Processing (ICASSP)}, 2018.

\bibitem{Nash1951}
J.~Nash, \emph{{Non-cooperative games}}.\hskip 1em plus 0.5em minus 0.4em\relax
  Annals of Mathematics, 1951.

\bibitem{dekerret2018_ISWCS}
P.~de~Kerret and D.~Gesbert, ``{Robust decentralized joint precoding using team
  deep neural network},'' in \emph{Proc. IEEE International Symposium on
  Wireless Communication Systems (ISWCS)}, Aug. 2018.

\bibitem{Kim2018}
M.~Kim, P.~de~Kerret, and D.~Gesbert, ``{Robust decentralized joint precoding
  using team deep neural network},'' in \emph{Proc. IEEE Asilomar Conference on
  Signals, Systems and Computers (ACSSC)}, Nov. 2018.

\bibitem{Lecun2012}
Y.~A. LeCun, L.~Bottou, G.~B. Orr, and K.-R. Muller, ``{Efficient BackProp},''
  in \emph{Neural Networks: Tricks of the Trade. Lecture Notes in Computer
  Science, vol 7700. Springer}, 2012.

\bibitem{Giaconi:SM:18}
G.~{Giaconi}, D.~{Gunduz}, and H.~V. {Poor}, ``Privacy-aware smart metering:
  Progress and challenges,'' \emph{IEEE Signal Processing Magazine}, vol.~35,
  no.~6, pp. 59--78, Nov 2018.

\bibitem{V2X:privacy}
N.~{Saxena}, S.~{Grijalva}, V.~{Chukwuka}, and A.~V. {Vasilakos}, ``Network
  security and privacy challenges in smart vehicle-to-grid,'' \emph{IEEE
  Wireless Communications}, vol.~24, no.~4, pp. 88--98, Aug 2017.

\bibitem{edge:bennis}
\BIBentryALTinterwordspacing
J.~Park, S.~Samarakoon, M.~Bennis, and M.~Debbah, ``Wireless network
  intelligence at the edge,'' \emph{CoRR}, vol. abs/1812.02858, 2018. [Online].
  Available: \url{http://arxiv.org/abs/1812.02858}
\BIBentrySTDinterwordspacing

\bibitem{FL:KonecnyMRR16}
\BIBentryALTinterwordspacing
J.~Konecn{\'{y}}, H.~B. McMahan, D.~Ramage, and P.~Richt{\'{a}}rik, ``Federated
  optimization: Distributed machine learning for on-device intelligence,''
  \emph{CoRR}, vol. abs/1610.02527, 2016. [Online]. Available:
  \url{http://arxiv.org/abs/1610.02527}
\BIBentrySTDinterwordspacing

\bibitem{DL_CommsEff}
M.~Li, D.~G. Andersen, A.~J. Smola, and K.~Yu, ``Communication efficient
  distributed machine learning with the parameter server,'' in \emph{Advances
  in Neural Information Processing Systems 27}, Z.~Ghahramani, M.~Welling,
  C.~Cortes, N.~D. Lawrence, and K.~Q. Weinberger, Eds.\hskip 1em plus 0.5em
  minus 0.4em\relax Curran Associates, Inc., 2014, pp. 19--27.

\bibitem{DCLimitedPrecisionGupta}
S.~Gupta, A.~Agrawal, K.~Gopalakrishnan, and P.~Narayanan, ``Deep learning with
  limited numerical precision,'' in \emph{ICML}, Jul. 2015.

\bibitem{DCOneBitQuan}
F.~Seide, H.~Fu, J.~Droppo, G.~Li, and D.~Yu, ``1-bit stochastic gradient
  descent and its application to data-parallel distributed training of speech
  {DNN}s,'' in \emph{INTERSPEECH}, Singapore, Sep. 2014, pp. 1058--1062.

\bibitem{ScalableDNNStorm}
N.~Strom, ``Scalable distributed {DNN} training using commodity gpu cloud
  computing,'' in \emph{INTERSPEECH}, 2015.

\bibitem{DCAjiSparse}
A.~F. Aji and K.~Heafield, ``Sparse communication for distributed gradient
  descent,'' \emph{arXiv:1704.05021v2 [cs.CL]}, Jul. 2017.

\bibitem{DCSattlerSparseBinary}
{F. Sattler et al.}, ``Sparse binary compression: {Towards} distributed deep
  learning with minimal communication,'' \emph{arXiv:1805.08768v1 [cs.LG]}.

\bibitem{McMahan2017CommunicationEfficientLO}
{H. B. McMahan et al.}, ``Communication-efficient learning of deep networks
  from decentralized data,'' in \emph{Proc. AISTATS}, 2017.

\bibitem{LAGGradientGiannakis}
{T. Chen et al.}, ``{LAG}: {Lazily} aggregated gradient for
  communication-efficient distributed learning,'' \emph{arXiv:1805.09965
  [stat.ML]}, May 2018.

\bibitem{Amiri:edge}
\BIBentryALTinterwordspacing
M.~M. Amiri and D.~G{\"{u}}nd{\"{u}}z, ``Machine learning at the wireless edge:
  Distributed stochastic gradient descent over-the-air,'' \emph{CoRR}, vol.
  abs/1901.00844, 2019. [Online]. Available:
  \url{http://arxiv.org/abs/1901.00844}
\BIBentrySTDinterwordspacing

\bibitem{Zhu:Huang:edge}
\BIBentryALTinterwordspacing
G.~Zhu, Y.~Wang, and K.~Huang, ``Low-latency broadband analog aggregation for
  federated edge learning,'' \emph{CoRR}, vol. abs/1812.11494, 2018. [Online].
  Available: \url{http://arxiv.org/abs/1812.11494}
\BIBentrySTDinterwordspacing

\bibitem{Amiri:fading:edge}
M.~M. Amiri and D.~G{\"{u}}nd{\"{u}}z, ``Over-the-air machine learning at the
  wireless edge,'' in \emph{submitteed, SPAWC}, 2019.

\bibitem{simeone:wirelessComputation}
\BIBentryALTinterwordspacing
S.~Ha, J.~Zhang, O.~Simeone, and J.~Kang, ``Coded federated computing in
  wireless networks with straggling devices and imperfect {CSI},'' \emph{CoRR},
  vol. abs/1901.05239, 2019. [Online]. Available:
  \url{http://arxiv.org/abs/1901.05239}
\BIBentrySTDinterwordspacing

\bibitem{Li:AI_5G:17}
R.~{Li}, Z.~{Zhao}, X.~{Zhou}, G.~{Ding}, Y.~{Chen}, Z.~{Wang}, and H.~{Zhang},
  ``Intelligent 5g: When cellular networks meet artificial intelligence,''
  \emph{IEEE Wireless Communications}, vol.~24, no.~5, pp. 175--183, October
  2017.

\bibitem{Kibria:BigData:5G}
\BIBentryALTinterwordspacing
M.~G. Kibria, K.~Nguyen, G.~P. Villardi, K.~Ishizu, and F.~Kojima, ``Big data
  analytics and artificial intelligence in next-generation wireless networks,''
  \emph{CoRR}, vol. abs/1711.10089, 2017. [Online]. Available:
  \url{http://arxiv.org/abs/1711.10089}
\BIBentrySTDinterwordspacing

\bibitem{Kafle:Slicing:18}
V.~P. {Kafle}, Y.~{Fukushima}, P.~{Martinez-Julia}, and T.~{Miyazawa},
  ``Consideration on automation of 5g network slicing with machine learning,''
  in \emph{2018 ITU Kaleidoscope: Machine Learning for a 5G Future (ITU K)},
  Nov 2018, pp. 1--8.

\bibitem{Venieris:DNN_Embedded:18}
\BIBentryALTinterwordspacing
S.~I. Venieris, A.~Kouris, and C.~Bouganis, ``Deploying deep neural networks in
  the embedded space,'' \emph{CoRR}, vol. abs/1806.08616, 2018. [Online].
  Available: \url{http://arxiv.org/abs/1806.08616}
\BIBentrySTDinterwordspacing

\bibitem{Yoo:AI_Silicon:19}
H.~{Yoo}, ``Intelligence on silicon: From deep-neural-network accelerators to
  brain mimicking {AI-SoCs},'' in \emph{IEEE Int'l Solid- State Circuits Conf.
  - (ISSCC)}, Feb 2019, pp. 20--26.

\end{thebibliography}

\end{document}